\documentclass[acmsmall]{acmart}

\usepackage{graphicx}
\usepackage{color}
\usepackage{xspace}
\usepackage{amsmath,amsfonts}
\usepackage{algorithm}
\usepackage{algorithmic}
\usepackage{textcomp}
\usepackage{xcolor}
\usepackage{booktabs}
\usepackage{booktabs,xltabular}
\usepackage{subcaption}

\newcommand{\TheName}{\textsc{EyeTrans}}

\newcommand{\squishlist}{
 \begin{list}{$\bullet$}
   { \setlength{\itemsep}{0pt}
     \setlength{\parsep}{0pt}
     \setlength{\topsep}{0pt}
     \setlength{\partopsep}{0pt}
     \setlength{\leftmargin}{2.5em}
     \setlength{\labelwidth}{1.5em}
     \setlength{\labelsep}{0.5em} } }

\def\BibTeX{{\rm B\kern-.05em{\sc i\kern-.025em b}\kern-.08em
    T\kern-.1667em\lower.7ex\hbox{E}\kern-.125emX}}

\newcommand{\squishend}{
  \end{list}  }

\usepackage{hyperref}
\hypersetup{
    colorlinks=true,
    linkcolor=green,
    citecolor={blue!50!black},
    urlcolor=green
}

\setcopyright{rightsretained}
\acmDOI{10.1145/3643732}
\acmYear{2024}
\copyrightyear{2024}
\acmSubmissionID{fse24main-p56-p}
\acmJournal{PACMSE}
\acmVolume{1}
\acmNumber{FSE}
\acmArticle{6}
\acmMonth{7}
\received{2023-09-28}
\received[accepted]{2024-01-23}

\begin{document}

\author{Yifan Zhang}
\orcid{0000-0001-5719-772X}
\affiliation{%
  \institution{Vanderbilt University}
  %\city{}
  \country{USA}
}
\email{yifan.zhang.2@vanderbilt.edu}

\author{Jiliang Li}
\orcid{0009-0004-2852-0865}
\affiliation{%
  \institution{Vanderbilt University}
  %\city{}
  \country{USA}
}
\email{jiliang.li@vanderbilt.edu}

\author{Zachary Karas}
\orcid{0000-0002-5721-8794}
\affiliation{%
  \institution{Vanderbilt University}
  %\city{}
  \country{USA}
}
\email{z.karas@vanderbilt.edu}

\author{Aakash Bansal}
\orcid{0000-0001-7475-7899}
\affiliation{%
  \institution{University of Notre Dame}
  %\city{}
  \country{USA}
}
\email{abansal1@nd.edu}

\author{Toby Jia-Jun Li}
\orcid{0000-0001-7902-7625}
\affiliation{%
  \institution{University of Notre Dame}
  %\city{}
  \country{USA}
}
\email{toby.j.li@nd.edu}

\author{Collin McMillan}
\orcid{0009-0005-0887-1083}
\affiliation{%
  \institution{University of Notre Dame}
  %\city{}
  \country{USA}
}
\email{cmc@nd.edu}

\author{Kevin Leach}
\orcid{0000-0002-4001-3442}
\affiliation{%
  \institution{Vanderbilt University}
  %\city{}
  \country{USA}
}
\email{kevin.leach@vanderbilt.edu}

\author{Yu Huang}
\orcid{0000-0003-2730-5077}
\affiliation{%
  \institution{Vanderbilt University}
  %\city{}
  \country{USA}
}
\email{yu.huang@Vanderbilt.Edu}

% \setcopyright{acmlicensed}
% \acmJournal{PACMSE}
% \acmYear{2024} \acmVolume{1} \acmNumber{FSE} \acmArticle{6} \acmMonth{7}\acmDOI{10.1145/3643732}

% \title{Incorporating Programmer Attention into Neural Code Summarization}
% \title{PGTrans: Programmer-Guided Transformer for Neural Code Summarization}
% \title{Eye-Tracking Transformer: Merging Human and Machine Attention for Neural Code Summarization}
\title{EyeTrans: Merging Human and Machine Attention for Neural Code Summarization}

% Placeholder - Traditional Transformer architectures have set new benchmarks in code summarization tasks but lack the human-centric focus that is critical for nuanced understanding. To bridge this gap, we introduce EyeTrans, an enhanced Transformer model that integrates human eye-tracking data collected through a Tobii eye-tracker. We validate EyeTrans under two experimental settings: Extreme Summarization and General Code Summarization. In both paradigms, EyeTrans significantly improves performance metrics—accuracy and F1 score for Extreme Summarization, and BLEU and GLEU scores for General Code Summarization—compared to the baseline Transformer model. Our results demonstrate EyeTrans's superior generalizability and robustness, making it a promising approach for the future development of Transformer-based models and Large Language Models (LLMs).

\begin{abstract}
% NOTSUREIFFIXED: \yu{currently it seems the motivation of using human atention is just "it is never been considered before". But I think it is also because "human attention proves to be important" and "it is never used before", thus you propose the first work to integrate human attention. maybe consider structure the abstract in this way: (1) as an representitive task of code comprehension, code summerization is so important and researchers have done a lot of work on it - both for understanding human attention and generating tools to automate it; (2) it has shown human attention indicate certain patterns and also relate to quality of summarization. (3) but so far those AI models have only used code features. (4) we hypothesize human attention can be leveraged to guide the model trainings which is completely overlooked before. (5) to test this hypothesis, look, eyetrans!}

%Neural code summarization leverages the progress of artificial intelligence to automatically summarize code snippets.
Neural code summarization leverages deep learning models to automatically generate brief natural language summaries of code snippets.
The development of Transformer models has led to extensive use of attention during model design.
%With the development of Transformer models, machine attention is extensively utilized in their model design.
%However, despite considering the static structure in source code, such as the Abstract Syntax Tree (AST), to encode structural information into attention, few studies consider human attention, specifically, programmer’s attention during manual summarization, which is typically more intuitive and has the potential to be synthesized into machine attention.
While existing work has primarily and almost exclusively focused on static properties of source code and related structural representations like the Abstract Syntax Tree (AST), few studies have considered human attention --- that is, where programmers focus while examining and comprehending code.
In this paper, we develop a method for incorporating human attention into machine attention to enhance neural code summarization.
% We argue that merely incorporating human attention into machine attention can enhance neural code summarization.
To facilitate this incorporation and vindicate this hypothesis, we introduce \TheName{}, which consists of three steps: (1) we conduct an extensive eye-tracking human study to collect and pre-analyze data for model training, (2) we devise a data-centric approach to integrate human attention with machine attention in the Transformer architecture, and (3) we conduct comprehensive experiments on two code summarization tasks to demonstrate the effectiveness of incorporating human attention into Transformers. 
Integrating human attention leads to an improvement of up to 29.91\% in Functional Summarization and up to 6.39\% in General Code Summarization performance, demonstrating the substantial benefits of this combination. 
We further explore performance in terms of robustness and efficiency by creating challenging summarization scenarios in which \TheName{} exhibits interesting properties. 
We also visualize the attention map to depict the simplifying effect of machine attention in the Transformer by incorporating human attention. This work has the potential to propel AI research in software engineering by introducing more human-centered approaches and data.

% FIXED: YU: not sure why you say "Transformers", you only have one vanilla transformer - Answer: usually all other Transformers adopt the Transformer block so I simplified a bit, but it should be changed to Transformer to prevent overclaim
% FIXED: YU: corresponding to CMK's comment: maybe we don't need to say "EyeTrans outperforms Transformer" but "integrating human attention leads to a X\% improvement on the performance" instead - and yes you may want to list a number here to indicate the improvement.
%FIXED: FIXME-kleach careful with "significantly" -- is it statistical significance? if not, consider "substantially" or "dramatically"

% Experimental results indicate that \TheName{} significantly outperforms the Transformer, showing a correlation between the quality of eye-tracking data and improvements in model performance, training efficiency, generalizability, and robustness.

%FIXED: FIXME-kleach suggest a "so what?" sentence.  We don't want to overclaim, but how do we think this work is going to influence software engineering in the future?

\end{abstract}

% \setcopyright{rightsretained}
% \acmJournal{PACMSE}
% \acmYear{2024} \acmVolume{1} \acmNumber{FSE} \acmArticle{6} \acmMonth{7}\acmDOI{10.1145/3643732}




\maketitle

\section{Introduction}\label{sec:intro}

% 1. Comprehension is a task that can represent how humans understand code
% EL: too early to bring in attention switch concept; 
A code summary is a brief description representing the purpose and function of code that can aid developer comprehension~\cite{stapleton2020human}.
% In code comprehension, programmers must formulate an understanding of code snippets based on the code content ~\cite{stapleton2020human}. % ZK - trying to think of alternatives for 2nd sentence here
%When a programmer reads a code snippet, they spend time and effort comprehending the code in question~\cite{stapleton2020human}. 
Code summaries are a common part of documenting source code --- while typically provided by the developer, machine learning models have increasingly been used to automatically generate summaries to augment documentation and improve comprehension. 
When documenting or summarizing source code, programmers' attention will focus on different parts of the code to formulate a general idea about it, eventually establishing a comprehensive understanding~\cite{haiduc2010supporting}.  %use~\cite not \cite

% FIXED: [whether it's code summarization or neural code summarization - it's a subtask of code summarization \yu{Prolly dont need to worry too much about the difference between code sum and neural code sum here}]

% 2. eye tracking captures patterns about code comprehension
% 3. some saccade indicate attention switching, which indicates (maybe how humans associate two tokens during comprehension?) see Section 3.2, 3.3 for relevant information.
% FIXED: \yu{state that "it is important to understand how developers do code sum because - tool design, - training, etc." then talk about the eye tracking studies}

%It is important to understand how developers do code summarization, because it involves software tool design, model training, and so forth. 
% this sentence doesn't make sense

%Studies abound regarding what actually occurs during this comprehension process.  

Several studies have developed an initial understanding of how human developers comprehend code.
To capture the nuanced attention shifts of programmers reading code, a popular approach is to conduct eye-tracking studies~\cite{abid2019using, rodeghero2015empirical} to empirically analyze common eye gaze patterns during the process. 
%Intuitively, certain %sac?
%eye movements
%can correspond to attention switching among different parts of the code when the human fixation position  %used before define?
%significantly changes from one concentrated area to another in the code snippet~\cite{sun2019smart, peitek2020drives}. 
Intuitively, certain patterns of eye gaze when examining source code correspond to a change in attention within the code~\cite{sun2019smart, peitek2020drives},
which is usually highly correlated with changes in human concentration and is potentially valuable for gaining a better understanding of the cognitive processes in programming~\cite{bansal2023modeling,sharafi2021toward, huang2020biases, ahmad2023we}. 
%This motivation has spurred advancements in human studies on eye-tracking patterns in programming. 
Researchers have conducted several eye-tracking studies to explore human attention during code summarization. 
These usually involve university students or professional developers drafting or evaluating summaries for code snippets and recording their eye gaze patterns to understand human attention in such activities~\cite{rodeghero2014improving,rodeghero2015eye}. Though the focus has been primarily on understanding the eye-tracking patterns, 
these studies have unveiled certain correlations between human visual attention and code summarization~\cite{rodeghero2015empirical, abid2019using}, 
% not sure how to phrase it yet- the previous one is fine 
paving the way for future work to improve relevant tool design leveraging human attention. 

% these studies have unveiled correlations between human visual attention and writing summarization. Such past works paved the way for future work to improve relevant tool design leveraging human attention. 
%So far, the focus has been primarily on investigating eye-tracking patterns to manual summarization or to traditional automatic summarization methods.

% UNFIXED: \yu{second Eric's comment: what nodes mean is unclear yet. You can start with eye tracking studies for code comprehension -- see my comment 6 as well}
% NOTSUREIFFIXED: EL: too early to bring in saccade probably 
% FIXED: \yu{this paragraph can be combined with the one above and shortened.}
% NOTSUREIFFIXED: ZK feels we need to motivate the human eye-tracking study some point earlier than contributions 
% EL: bring in the concept of automatic summarization wihtou introducing it first

% UNFIXED: \yu{I would adjust the order of the content in the following part on AI models for code sum.  you say AImodels are developed for code sum -> transformers are used and it is one of the most advanced model-> attention in transformer -> }
In recent years, in tandem with the traditional definition of ``attention'' in human activities, \textit{attention} can also refer to a concept in machine learning that describes the importance of weights in the layers of neural networks~\cite{niu2021review, guo2022attention}. 
Within this area, the \textit{Transformer}~\cite{vaswani2017attention} architecture is one of the most influential baseline structures that incorporate \emph{neural attention} into a machine learning model. 
With advancements in machine attention mechanisms for system applications~\cite{zhao2022queryformer, wang2022jtrans, zhu2023ktrans}, an increasing number of Transformer-based models have been developed to address a variety of neural code summarization tasks~\cite{tang2022ast, ahmad2020transformer, gong2022source, tang2021ast}. These methods leverage specific domain knowledge to modify their model structures accordingly, incorporating elements such as the Abstract Syntax Tree (AST)~\cite{tang2022ast}, dataflow graph~\cite{gong2022source}, and function call graph~\cite{le2022autopruner}. 
% UNFIXED \yu{I would say something like "while human attention implies potentials to indicate critical information in source code for comprehension and transformer-based model aims to develop attention map, we missed the opportunity of combining them/it has never been done/none of the current work has tried..." I prolly won't use the term "dynamic human attention" -- Idont think there are such terms. just say "human attention" or human visual attention is fine}
To generate code summaries, these Transformer models essentially aim to ``comprehend'' the code, which is a task that human programmers already accomplish. 
It is therefore possible that human attention can be leveraged to improve these models' performance~\cite{rohanian2017using}, which, to the best of our knowledge, has not yet been attempted. 
This process can be framed into the following questions: can human attention be integrated into complex machine attention mechanisms to enhance overall performance? If so, how can we pragmatically incorporate human attention into attention-based machine learning models?
% However, none of these models incorporate human attention, opting instead to integrate static structures that are already part of the implicit information within the code. Incorporating human attention is presumably more intuitive and can complement the training processes of Transformer models. 
% UNFIXED % \yu{I like this paragraph, but it does not flow well from the paragraph above-- also can be shortened with the last part of the paragraph above.} This leaves one significant question largely under-explored by many researchers in both the software engineering and AI communities: can human attention be integrated into complex machine attention mechanisms to enhance overall performance, and if so, how can we pragmatically incorporate human attention into attention-based machine learning models? 
These questions are particularly relevant for Software Engineering (SE) and AI communities considering the Transformer architecture, which underlies the most recent advanced AI applications in SE, and large language models (LLMs) in natural language processing (NLP)~\cite{liu2019roberta,guo2020graphcodebert,touvron2023llama}.

% FIXED: \yu{it is a bit weird to me that Transformer is capitalized. It is not a speficific model, I dont think we need to capitalize it} - Answer: usually non-capitalized transformer means the real transformer in vehicles or hardwares. Capitalized Transformer means the model in AI

% 4. Thus the problem is whether we can use human attention patterns (i.e. attention switch) to improve ...

\begin{figure}[t]
\centering
\includegraphics[width=.7\textwidth]{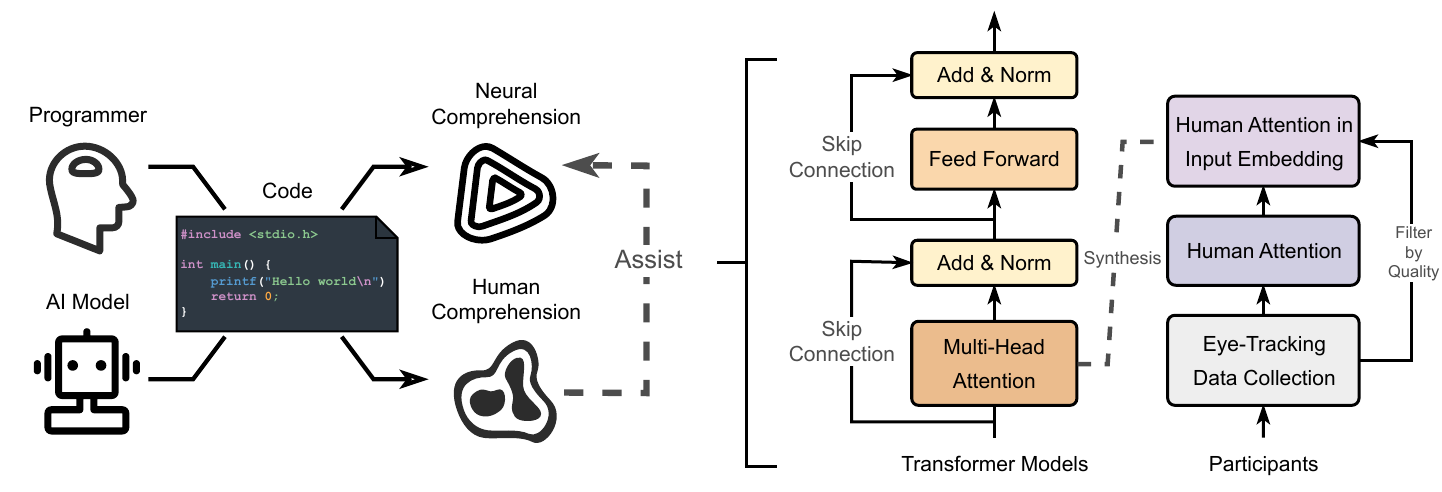}
\caption{Conceptual Overview of \TheName{}. \TheName{} synthesizes human attention information into the self-attention mechanism of Transformers, assisting Transformers in code summarization tasks.}
\label{fig:intro}
\end{figure}

% FIXED \yu{I would not say "driven bu intuition". Instead you can say "we hypothesize ..."}
In this paper, we hypothesize that human attention can enhance Transformer models, and with the understanding that a formalized methodology for this integration has yet to be explored, we introduce \TheName{}. \TheName{} leverages human eye-tracking data during training to effectively integrate human attention with machine attention, as illustrated in Figure~\ref{fig:intro}. 
% During inference, no eye-tracking information is needed. % CRFIX
To this end, we first conduct a human study to collect eye-tracking data from 27 programmers, which we then preprocess and incorporate into the Transformer. We describe our methodology, then present evaluation metrics that assess \TheName{}'s performance. 
Specifically, we employ two code summarization-based evaluation methods to illustrate the effectiveness of human attention in augmenting neural code summarization:
%To represent the initial phase of code comprehension, where a programmer aims to categorize code into broader genres or functional categories to acquire an overview of the code, we use 
(1) classification-based code summarization, referred to as \textit{Functional Summarization}, in which a model predicts the name of function (e.g., ``open'' or ``sort'') given its source code. This serves as a foundational step, enabling programmers to consolidate a preliminary understanding of the inherent functionality and semantics of the code. 
%After gaining a foundational overview, we then use 
(2) Sequence-to-sequence (seq2seq) based code summarization, referred to as \textit{General Code Summarization}, in which a model generates a sequence of natural language tokens succinctly describing what a given function does (e.g., documentation for the function).
%which emulates the advanced stage of code summarization, where programmers extract more specific and detailed information from a code snippet. 
%This advanced technique facilitates a deeper and more nuanced understanding of the code’s structure and functionality, allowing for the extraction of intricate details and subtleties of the code’s implementation. 
We find that integrating human attention results in a performance improvement of up to 29.91\% in functional summarization, and up to 6.39\% in general code summarization, thereby demonstrating the effectiveness of \TheName{}. 
We view our work as a proof-of-concept and hope our work can inspire more future work to leverage human factor research in AI for software engineering. 
We included the eye-tracking data collected in the study (de-identified) and the experimental code in the supplementary package and plan to release them to the public. % \yu{FIXME}.
% [FIXME Main results here]

% FIXED \yu{here you should provide a high level intro of your propsoed work: first conduct an eye trackign study with X participants -> analyze the integrate the visual attention patterns in transformer -> evaluation -> results + numbers }[we collect a eye-tracking dataset for this purpose] 
The contributions of this paper are as follows:
\squishlist
    \item We propose \TheName{}, the first approach to demonstrate that integrating human attention with the Transformer architecture can substantially enhance the performance of neural code summarization.
    % FIXME do not use significant unless you have statistical significance.
    \item We conduct exploratory examinations of eye-tracking data and rationalize the integration of human attention through preliminary analysis.
    \item We design a data-centric method to incorporate human attention into Transformers without altering its structure, and present comprehensive quantitative analyses to validate the performance of our approach.
    % Kevin: Emphasize the importance of not altering Transformer structure
    \item We illustrate alterations in the Transformer's attention maps to qualitatively demonstrate the changes brought by combining human attention and attention layer in Transformer.
    \item We review existing attention synthesis methods in relation to our work and outline prospective avenues for future research.
\squishend

% UNFIXED: \yu{this part can be shortedned and moved to where comments 13 is, leaving contribution as the last part of intro.} -- I fixed it - yu

% FIXED: \yu{you should list some numbers to show your improvement -- is your hypothesis true? Does Eyetrans work?} -> Answer: listed the percent of improvement 

% Kevin: Cite something about extreme code summarization

% Kevin: We show human attention improves in xxx way, and can generalize in many Transformer xxx, also mention some numbers here

\textbf{Paper Organization:} 
%The remainder of this paper is organized as follows. 
Section~\ref{sec:preliminary} presents the design for the eye-tracking human study in this paper, and provides background knowledge on \TheName{} and the Transformer architecture. Section~\ref{sec:approach} elaborates on the details of \TheName{}. Section~\ref{sec:experiment} details the experimental setup. Section~\ref{sec:analysis} analyzes the results. Section~\ref{sec:validity} discusses threats to validity. Section~\ref{sec:background} surveys related work. Finally, Section~\ref{sec:conclusion} concludes the paper and outlines future research directions.

\section{Preliminaries}\label{sec:preliminary}
%Yu: this begining part reads a bit weird
The high-level idea of \TheName{} is to integrate human attention, collected from eye-tracking, into Transformer models to enhance the performance on code summarization.
In this section, we first introduce the design for the human eye-tracking study, where the \textbf{eye-tracking data} is collected and used in \TheName{}, and then provide relevant background information for the \textbf{Transformer} architecture, especially the self-attention module of the Transformer model. 
%We collected eye-tracking data from programmers as they examine and summarize snippets of source code.  In turn, this eye tracking data is used to improve a AI-based summarization model by capturing what humans focus on when they comprehend source code.
% the first sentence above was added after the following sentences were written - but the transition seems a bit off maybe?
%Given that our approach is contingent upon human eye-tracking and the Transformer NLP architecture, we 
 %for the collection process of \textbf{eye-tracking data}, including the task materials, recruitment, and experimental protocol. We also provide an introduction to the Transformer architecture, especially the self-attention module of the \textbf{Transformer} model.

% FIXED: \yu{+ especially focusing on the attention structure in transformers -- people already know transformers, you need to point out here you emphasize on the the attention part that is more relevant to your model}. -> slightly changed the sentence

\subsection{Eye-Tracking Study Design}\label{sec:eye-tracking}
We first conducted an eye-tracking study with 29 student programmers where participants completed Java source code summarization tasks. 
%In this task, participants read snippets of Java code and wrote short, corresponding descriptions in natural language \cite{zhang2022survey}. 
In this subsection, we describe details of the task design and experiment protocol of this human study.
%Java methods and their integration into the task, as well as the eye-tracking data collection and protocol.

\begin{figure*}[tb]
\centering
\includegraphics[width=0.78\textwidth]{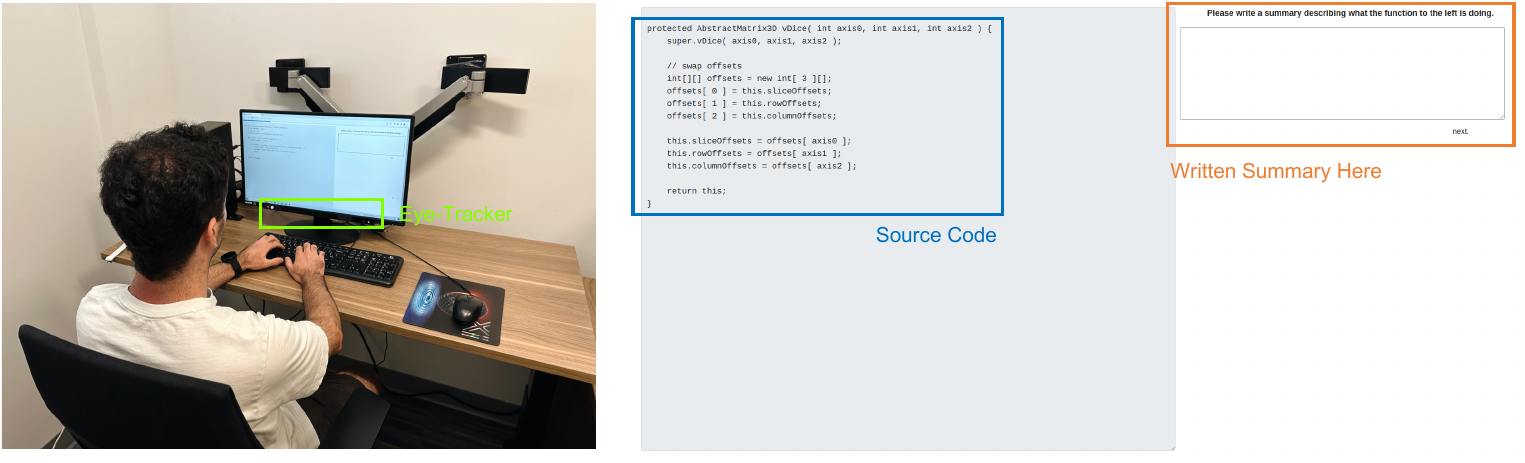}
\caption{(\textit{Left}) The experimental room, with the task displayed on the monitor. The Tobii Pro Fusion Eye-tracker is a thin bar magnetized to a strip at the bottom of the monitor. (\textit{Right}) A screenshot of one task example, with the Java method displayed on the left of the screen, and the summary writing location in the top right.}
\label{fig:reading_writing}
\end{figure*}

\textbf{Java Methods and Task Design} 
% \yu{this part can be shortened. too wordy.,. I will give you one example here: "The Java methods used in this study originate from the publicly available FunCom dataset \cite{leclair2019recommendations}. The dataset contains roughly 2.1 million Java methods, but we filtered the corpus based on recommend quality standards to 190,000 \cite{allamanis2017learning}. From here, a random subset of methods was selected to present during the task, which was filtered down to 162 based on Java readability standards \cite{network1999code}."  -> "We constructed a set of 162 java methods for summerization tasks in our study from the FunCom dataset[cite]. These methods were randomly selected, but filtered based on their length (to fit on the monitor without scrolling).   " then you introduce finally what these 162 stimuli looks like, complexity, LOC, etc.}
We constructed a dataset of 162 Java methods for summarization tasks in our study using the FunCom dataset~\cite{leclair2019recommendations}. 
These methods were randomly selected from FunCom, then filtered based on their length to fit on the monitor without scrolling\footnote{A static screen of code that does not scroll facilitates the collection of eye tracking data.}.
% The Java methods used in this study originate from the publicly available FunCom dataset \cite{leclair2019recommendations}. 
% The dataset contains roughly 2.1 million Java methods, but we filtered the corpus based on recommend quality standards to 190,000 \cite{allamanis2017learning}. 
% From here, a random subset of methods was selected to present during the task, which was filtered down to 162 based on Java readability standards \cite{network1999code}.
% of 205 methods was randomly selected to present to participants during the experiment. Some methods contained long lines of code that wrapped onto the next line. For readability, we excluded these methods, yielding 162 methods in the final dataset.
In this final dataset of 162 Java methods, the shortest contained 5 lines of code, while the longest contained 26 ($\mu$=11.72, $\sigma$=4.25). 
The most complicated method had a cyclomatic complexity of 11 and
%meaning it contained 11 linearly independent paths. 
the simplest had a complexity of 1 ($\mu$=2.59, $\sigma$=1.56). 
In the experiment, methods were presented one at a time, along with an empty text box for participants to type their summaries, as shown in Figure~\ref{fig:reading_writing}.
% \yu{wordy wordy....just one sentence please "the methods were presented on a 24" monitor following standard Java format without syntax highlighting." then you move the eye tracker part here say "participants' eye gaze data was recorded during the tasks using the the Tobii Pro Fusion eye-tracker (60Hz), which is accurate down to 0.1--0.2 inches (0.26--0.53cm) on the monitor \cite{Tobii_2023}. "}
The methods were presented on a 24" monitor, without syntax highlighting and following standard Java formatting~\cite{network1999code}. Participants' eye gaze data was recorded during the tasks using the the Tobii Pro Fusion eye-tracker (60Hz), which is accurate to 0.1--0.2 inches (0.26--0.53cm) on the monitor~\cite{Tobii_2023}. 
% Using a custom web task, we displayed randomly selected methods from our dataset and recorded participant input on a 24" monitor.
%, with a refresh rate of 60Hz. 
% We indented and formatted these methods according to Java standards \cite{network1999code}, and displayed them at font size 14, without syntax highlighting.

\textbf{Recruitment and Experimental Protocol}  
We recruited 29 undergraduate and graduate CS students in our study with IRB approval at \textit{elided for double blind}. 
All participants had taken the Data Structures course or equivalent and passed a Java coding test during pre-screening. 
%Programmers 
% interested in the study would first email the researchers, then 
%completed the consenting and pre-screening processes electronically. 
%Programmers were eligible to participate if they were over 18 years of age, had at least one year of Java experience, had taken Data Structures, and had no history of epileptic seizures \cite{Tobii_2023}. We tested participants' basic understanding of Java with a prescreening question asking them to describe the purpose of an obfuscated Java method.
%, which was an in-order tree traversal. 
%comments for zach: those details are really not that necessary
All participants who completed the study were compensated \$60.
Due to a protocol error in one case and software malfunction in another, 27 of the 29 participants' data were included in the final dataset. These 27 participants were 23.8 years old on average, and include 8 women and 15 graduate students.

The experiment was conducted in person, in an office with natural lighting (as shown in Figure~\ref{fig:reading_writing}). During the entire experiment, participants completed 24 or 25 \textit{stimuli}, 
% \yu{why roughly??} 
where one stimulus consisted of a Java method and participants were asked to type their summaries for the method. %\footnote{A defect in stimulus presentation resulted in some participants receiving 24 methods and others receiving 25 methods to summarize}. 
%The discrepancy in stimulus number between participants is due to a small defect in stimulus presentation, which was patched. 
A break period was built halfway into the task interface, both for participants to rest and for the researcher to recalibrate the eye-tracker (for data quality).
% Finally, across all participants, we collected eye-tracking data for 661 summarization tasks using 67 unique Java methods. (\yu{FIXME unique Java methods}).
%Prior to the experimental task, participants completed a demographics questionnaire. 
Each participant took roughly 50 minutes to complete the entire experiment. 
%The task duration for each participant writing code summaries and completing the questionnaire was roughly 50 minutes.
%Eye-tracking data was collected using a Tobii Pro Fusion eye-tracker (60Hz), which is accurate down to 0.1--0.2 inches (0.26--0.53cm) on the monitor \cite{Tobii_2023}. 
%To record the eye-tracking data during the task, we integrated the Tobii Pro Software Developer Kit into the custom interface described above \cite{python_tobii_sdk_reference_guide}. 

\subsection{Self-Attention in Transformers}\label{sec:prelim:transformer} 
% \yu{change the section title to something like "self-attention in transformers" or "transformer and its self-attention architecture"}

% Kevin: make people realize that Transform is mainly on attention, good choice for our paper, also may need to clarify "though, this attention is slightly different than human attention", heat map to tell reviewers that which part of the model is important during training -> we influence this by introducing human attention (distinguish between human attention and machine attention)

% Yu: i.e., how we inspired by leveraging human attention to help enhance or guide attention

% Yu: Start a sentence "Informally, we hypothesize xxx"

% Kevin: Explanation about why is model attetnion not good enough, what you think human can help with -> intuition about current attention Transformer models, the attention layer learns "something specific that human cannot pay attention to"

The Transformer architecture~\cite{vaswani2017attention} is a seminal model in NLP and the backbone of \TheName{}. The architecture is highly reliant the \emph{self-attention mechanism}, 
which interrelates tokens in the input sequence with surrounding context. 
In this section, we highlight how  self-attention functions within the Transformer. 
Later, in Section~\ref{approach:modeling}, we discuss how we augment model self-attention with human attention to improve the Transformer's performance. 
%detailed illustration in Section~\ref{approach:modeling} about how our proposed integration of human attention is well-suited to enhance Transformer's performance.

Before input sequences are fed into the Transformer, each token in the sequence is mapped into an embedding space, where it is represented as a vector. 
% ZK - quick definition of embedding space
Conventionally, a token's input embedding is derived from the summation of its semantic and positional embeddings. 
In \TheName{}, we introduce a third embedding component. This additional embedding term relates pairs of tokens examined by human programmers in close temporal proximity, thus capturing human visual attention patterns during code comprehension.

%In Transformer architectures, self-attention is applied once the input embeddings are calculated.
In Transformer architectures, \emph{self-attention} helps the model learn how related two tokens' embedding representations are. 
Informally, the more related two tokens' embedding representations are, the more a self-attention layer learns to associate these tokens together. As such, Transformers' attention largely refers to an ability to learn the inter-relation between tokens. Consequently, we intuit that we can improve this self-attention mechanism by incorporating human attention based on the program elements humans consider during comprehension~\cite{ahmad2023we, ali2015empirical}. 
\emph{We hypothesize that a human's way of associating elements in code may capture subtle inter-relations that are not easily learned by the Transformer.} % Yu: love it love it! whoever wrote this sentence :) 

% Given the input embedding, the self-attention mechanism is then applied. 
More formally, the self-attention mechanism transforms each token's embedding into three vectors: query $(Q)$, key $(K)$, and value $(V)$. 
Conceptually, the key/value/query concept is analogous to a search engine. Each input token, denoted $token_0$, aims to find other related tokens. To accomplish this, $token_0$ uses its query vector, $Q_0$, to identify related tokens, similar to a search query into an indexed database. In response to this query vector, every other input token, denoted $token_i$, provides a key vector, $K_i$. If $Q_0$ closely matches some $K_i$, then the mechanism deems $token_i$ relevant to $token_0$, and presents $token_i$'s content in response, represented by its value vector, $V_i$. This vector $V_i$ is then mixed with $token_0$'s value vector $V_0$. 

Given input embedding matrix $X$, the matrices for query, key, and value are $Q=XW_q, K =XW_k, V = XW_v$, where $W_q, W_k, W_v$ are learned parameters. Self-attention is then calculated as follows:
\begin{equation}\label{eq:att}
    Attention(Q, K, V) = softmax(\frac{QK^T}{\sqrt{d_k}})V, \text{where} ~d_k\text{ is the dimension of key vector.}
\end{equation}
The $QK^T$ term in this equation is intuitively analogous to a similarity matrix representing pair-wise relatedness between tokens. 
%A heatmap of the $QK^T$ matrix, as we use in Section~\ref{sec:rq4}, is thus an conceptual representation the inter-relation of input tokens deemed by self-attention. %why do we care about heat maps here?

% [Transformers associate words that are semantically similar. Nevertheless, in code summarization tasks, a variable declaration may be deeply associated with the function declaration - and such an association is more difficult capture through embedding similaries alone. - thus human attention patterns associating tokens may be helpful]

Equation (\ref{eq:att}) describes a single instance of attention, but
in practice, multiple attention functions are performed in parallel, leading to multi-head attention. While attention is the essence of the Transformer, the architecture has other components that contribute to its effectiveness. As illustrated in Figure~\ref{fig:intro}, the multi-head attention's output is subject to layer normalization, fed through a fully connected network, and normalized again.  This sequence of layers, starting from multi-head attention to layer normalization, constitutes one \emph{Transformer block}.  When multiple blocks are connected in sequence, the output from one block serves as the input for the next, and the final output can be used for a variety of downstream tasks. In \TheName{}, we  employ stacks of Transformer blocks as the NLP model for code summarization, as discussed later in Section~\ref{sec:experiment_transformer}.
%FIXED need a "who cares" sentence... why do we need to know this? - EL: fixed withe last sentence
% : $MultiHead(Q, K, V ) = Concat(head_1, \cdots, head_h)W^O$, where $head_i = Attention(QW^Q_i, KW^K_i, VW^V_i)$ and $W^Q_i, W^K_i, W^V_i, W^O$ are matrix parameters.

% Intuitively, the Transformer self-attention's proclivity to  intuitive representation of attention and relatedness inspires our methodology to integrate human attention cues into machine attention, as illustrated in Section~\ref{approach:modeling}.

% The output for a transformer block is the result from the last normalization layer.

\section{Approach}\label{sec:approach}

In this section, we introduce the methodology in \TheName{} to combine machine and human attention to enhance neural code summarization.  
%In \TheName{}, the Transformer's self-attention constitutes the machine attention.  
We integrate human attention into \TheName{} by implanting eye-tracking data into the Transformer's input embeddings. 
Our approach for combining machine and human attention consists of the following steps: 
we (1) preprocess source code and (2) preprocess eye-tracking data to obtain preliminary datasets. 
Then, we (3) map the eye-tracking data onto the source code by combining their respective embedding representations.
In doing so, we can improve Transformer-based summarization by leveraging how humans examine and comprehend code when summarizing it.

% In \TheName{}, the integration of human and machine attention serves to enhance the model's performance in neural code summarization tasks. We now present each step of our approach.

\subsection{Source Code Pre-Processing}\label{src_code_preprocess}
To preprocess the source code, 
% preprocessing encompasses two steps: (1) converting source code into Abstract Syntax Tree (AST) representations and (2) augmenting ASTs to expand the dataset.
% \subsubsection{\textbf{Transformation of Source Code to AST}}
we first use the srcML parser to convert the textual format of each Java method into a corresponding AST representation \cite{collard2013srcml}. 
Transforming the code into ASTs is suitable for neural code summarization because this representation provides domain-specific, structural information of source code~\cite{alon2018code2seq, hu2018deep, leclair2020improved, shi2021cast, lin2021improving, tang2022ast, gong2022source, gao2023code}. 
% This transformation is beneficial as ASTs explicitly provide domain-specific, structural insights into the source code, a feature pivotal for neural code summarization. 
% EL: since we are defining literals a little differently in the next paragraph, the phrase "literals in `\texttt{i < 0}'" could be a little misleading
% ZK - makes sense
This structural information is encoded within the tree structure of nodes, where each node describes both token types and token literals. For instance, each of the three nodes in \{\texttt{`i', `<', `0'}\} may be a child node to a \texttt{conditional statement} node. 
% For instance, a \texttt{conditional statement} node comprised of `\texttt{i < 0}' has three child nodes: `i', `<',  `0'. 
The ASTs serve two purposes in this study: (1) to provide a suitable, structured representation for training a Transformer-based model, and (2) to facilitate mapping eye tracking data to corresponding program structures in a snippet of source code.

% EL: It's kind of important that we emphasize the memorization issue only becomes significant when the dataset is small (when trained on sufficeint Java methods, we no longer need to eliminate literls). Maybe we can move this first sentence to the end of the paragraph since it seems more fitting with the limited dataset size problem?
% ZK - in my discussions with Yu, she mentioned it's important to say *why* you're doing something. I reorganized it for that, but I'll fix the logic of it

\begin{figure*}[t]
\centering
\includegraphics[width=0.88\textwidth]{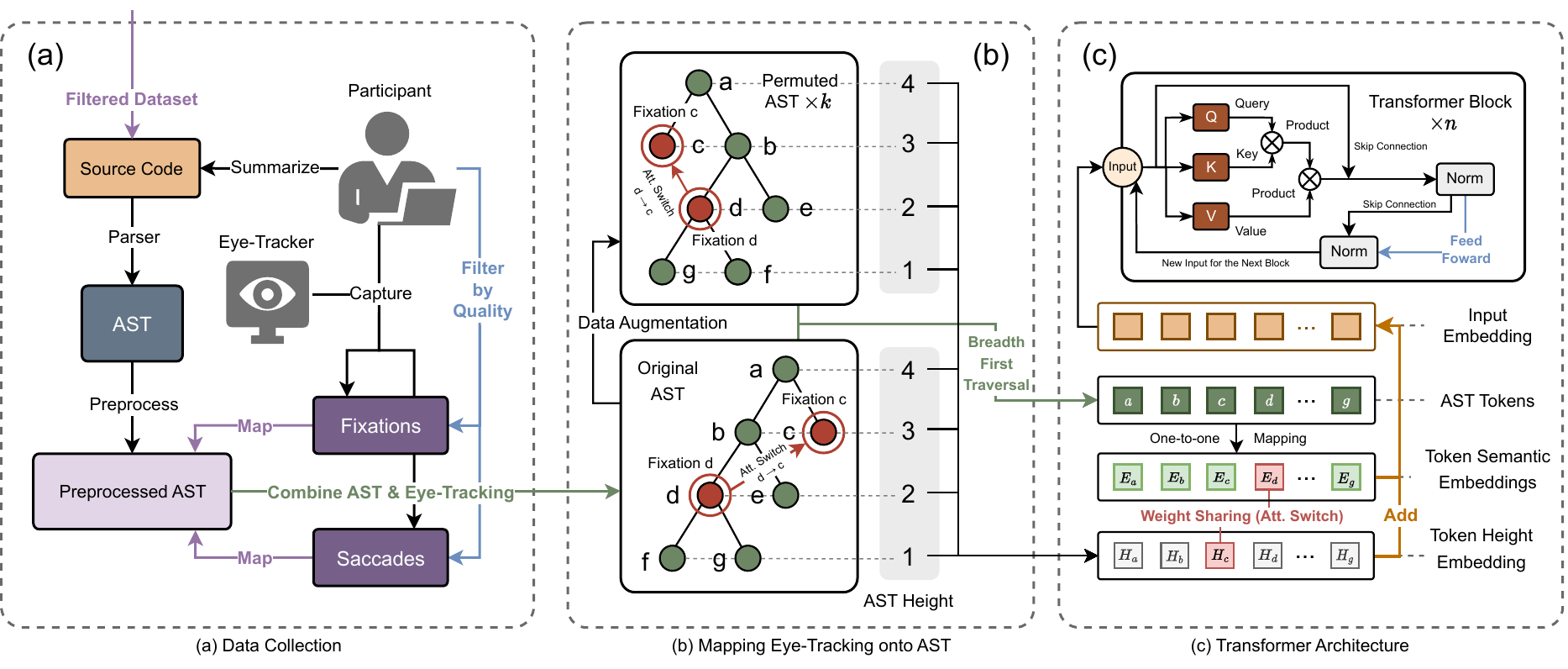}
\caption{\TheName{} Methodology Overview. (a) depicts our data collection processes. (b) describes the data preprocessing steps, including transformation to ASTs, data augmentation, and representing attention switches as edges on ASTs. (c) underlines our proposed approach to model attention switches using Transformers' default input embedding modality.}
\label{fig:model}
\end{figure*}

\textbf{Training Transformers on Code Structure}
We first parse source code into corresponding ASTs, and then discard token literals to abstract away individual program details. 
For example, a conditional expression such as \texttt{i < 0} may instead be represented as \emph{variable < integer} to avoid overfitting to specific token values while retaining appropriate structural program information. 
%To train Transformer-based code summarization models on AST representations of source code, we discarded the token literal names for reasons related to the small size of the dataset. 
%While 67 Java methods with corresponding eye-tracking data can be considered extensive from a human-study perspective, this is relatively sparse for NLP tasks. 
%FIXME by the way, this is the first mention of "67"  It was 162 above.   Which is it?
%This means that within such a dataset, each Java method likely contains literals from non-overlapping token spaces, which may cause overfitting in model training. 
%In other words, the NLP model may form associations between specific literals and the anticipated output, leading to undesirable memorization issues. 
%We therefore retain only the token types. 
%All ASTs within our dataset share a common set of token types. Thus, by extracting and using the structural information, we convey to the Transformer the structural essence of the code. 
%For example, the AST node for a Java variable named \texttt{foo} would retain the token type "name" while discarding the literal value "foo".
This approach aligns with the code structural learning settings investigated and validated by CodeNN~\cite{iyer2016summarizing}, emphasizing learning through code tokens devoid of literals.

%Beyond the elimination of token literals, we further address the constraints of dataset size by using \emph{data augmentation} to grow the dataset. 
We further employ \textbf{\emph{data augmentation}} to grow the dataset. 
Typical human eye-tracking studies do not yield the volume of data required for modern machine learning. Thus we develop the augmentation method below to overcome this disadvantage.   
Given the original AST of a Java method, we generate multiple variants of the AST (i.e., we augment the AST) via subtree permutations. Recursively, for each node within the AST, we randomly permute all subtrees of the node, altering the sequence of AST subtrees but preserving the hierarchical order (i.e., the parent-child relationship). 
For instance, in Figure~\ref{fig:model} (b), the augmentation alters the order of nodes `b' and `c', but `a' remains the parent of `b' and `c'. 
Although such permutations introduce changes to the semantics of the original program, they still preserve much of the core information of the original source code. 
For example, switching the order of two parallel assignment statements in a Java method generally retains the overall semantics. 
We empirically verify that the benefits of growing the dataset through augmentation outweigh the semantic deviations induced by AST permutations.

\textbf{Analyzing Eye-tracking Data using Code Structure} 
To analyze the eye-tracking data, we use the AST information to examine programmers' code reading patterns as they summarize code. 
Specifically, we sought to determine whether there exists consistent code-reading strategies that programmers used during the task. 
Informally, if programmers looked equally at every token literal in a snippet of code, it may not be meaningful to use eye-tracking data to assist neural code summarization.  
%To this end, we examine the most common transitions programmers made in their attention as they read code to understand the avenues of their attention. 
Thus, within the eye tracking data, we consider the most common transitions where the eye focuses within snippets of source code to represent programmers' attention.
For instance, if a programmer looks at a method declaration, the \emph{next} element they look at could be the return statement, other method calls, or the arguments of that method. 
%For instance, do programmers commonly look from the method declaration to return statements? 
%Or from method calls to arguments? 
Here, we map the eye-tracking data back to \emph{semantic categories} of structural program elements in the AST. 

To determine which semantic categories to consider, we referred to a widely-used Java textbook~\cite{schildt2007java}. 
The list of semantic categories includes the method declaration, variable declarations, return statements, conditional statements, and conditional blocks, among others. The complete list of 19 semantic categories can be found in the Supplementary Materials.
Multiple labels may apply to a single token literal. 
In these scenarios, we assigned the most meaningful labels, based on the semantics of Java, the structural context of the literal, and prior code summarization and comprehension research~\cite{rodeghero2014improving, p-Bednarik-strategy-2012, bergeretti1985information}.
For example, the equal sign is an operator, but can also belong to a variable declaration. 
In this case, we would label the equal sign as a `variable declaration,' because this label conveys more substantial information than `operator' in this context~\cite{borger1999programmer}. 
In brief, we extract the structural information of a program into an AST, then analyze the reading patterns of programmers as they comprehend code, and then map this information back to the AST in terms of high-level semantic categories.
%To summarize, we extract the structural information from the AST both to overcome limitations of the dataset size and to analyze the semantics of programmers' reading patterns as they comprehend code. 

\subsection{Eye-Tracking Data and Preprocessing}\label{sec:preprocess}
In essence, raw eye-tracking data consists of screen coordinates and their timestamps. 
To map these gaze coordinates to token literals on the screen, we calculated bounding boxes around each token using the \texttt{opencv-python} library for computer vision. Using these pixel-coordinate boundaries, we then localized gaze coordinates in the eye-tracking data to corresponding bounding boxes. 

Localizing gaze to token literals is a crucial preliminary step, but further processing is needed to make inferences about human cognition.  
Specifically, we extract characteristic visual patterns in humans' gaze behaviors, \emph{fixations} and \emph{saccades}, from the raw eye-tracking data~\cite{sharafi2020practical}. A fixation is a spatially-stable eye gaze which typically lasts for 200--300 milliseconds. Most cognitive processing of visual information occurs during fixations~\cite{p-sharafi-metrics-2015}.  Saccades are shorter in duration (40--50ms), and occur when humans make jumps in their visual field. In contrast to fixations, there is little cognitive processing that occurs during saccades~\cite{just1980theory}. 
Therefore, to understand patterns of human cognition, we needed to differentiate fixations from saccades~\cite{sharafi2020practical}. 
By current standards, this distinction is made using the \textit{velocity} of the eye-movement. In other words, if the speed of an eye-movement exceeds a certain threshold, a Velocity-Threshold Identification (I-VT) algorithm will identify it as a saccade~\cite{olsen2012tobii}. 
Based on our implementation and following best practices~\cite{birawo2022review}, we considered an eye-movement as a saccade if it exceeded 400px/100ms.

Once fixations are identified in the eye-tracking data, researchers typically use the number of fixations (i.e., \emph{fixation count}), and their duration (i.e., \emph{fixation duration}) as a proxy to measure cognitive processing \cite{p-sharafi-metrics-2015}. For instance, a higher fixation count and longer fixation durations signify greater cognitive effort~\cite{just1980theory}. 
By using fixations as building blocks, researchers can glean more complex cognitive behaviors from eye-tracking data. 
For instance, researchers can examine the \textit{ordered sequences} of fixations (i.e., scan paths) for insights into the flow of human attention~\cite{sharafi2020practical}. 
In this paper,  we use \textbf{attention switch} to refer to transitions between fixations within these ordered sequences (i.e., a saccade connecting two distant fixation points). 
% For instance, researchers can pre-define \textit{Areas Of Interest} (AOIs) in a visual stimulus, and measure participants' fixations in relation to these \cite{sharafi2020practical}. For example, Yusuf et al. measured programmers' fixations with respect to UML class diagrams, where each AOI was a component in the diagram. 
% FIXED \yu{attention switch is something you use in this paper, it does not mean "looks between AOIs". you can say "we use attention swtich to refer to the attention moving between fixations (i.e, scan path)."} \textbf{Attention Switch} Researchers also compute \emph{attention switches}, or the number of times a participant looks between AOIs, to measure how participants synthesize information between different sources \cite{ahmad2023we, ali2015empirical}. A higher number of attention switches also indicates a higher cognitive load \cite{sharafi2020practical}.
% In this study, we denote attention switches as transitions between tokens literals in code, (i.e., a saccade connecting two distant fixation points).
As such, each programmer's eye-tracking data on each Java method can be seen as a sequence of attention switches. 
% For the eye-tracking data, we explore code reading patterns by considering transitions between semantic categories (Section~\ref{src_code_preprocess}). 
% 

% EL wrote below
Each attention switch can also be represented on the AST as a directed edge from one node to another, as depicted in Figure~\ref{fig:model} (b). 
For each AST in our dataset, we therefore have derived attention switch edges from the human eye-tracking data.
% Naturally, each original AST is associated with a sequence of attention switch edges. 
When we augment the original ASTs via permutation to generate variants, we also map the attention switch edges onto the permuted ASTs, preserving the endpoints of each attention switch. For example, in Figure~\ref{fig:model} (b), an attention switch edge connecting nodes `d' an `c' in the original AST would still connect these two nodes in the permuted AST. Thus, every permuted AST possesses an enduring set of attention switch edges, which ensures the integrity of the attentional data amidst structural alterations. 

\subsection{Modeling Human Attention into Transformers}\label{approach:modeling}

In this section, we introduce our approach to model human attention into Transformer.  Our key idea centers around mapping \emph{attention switches}, defined in Section~\ref{sec:preprocess}, onto each of the corresponding augmented ASTs generated in Section~\ref{src_code_preprocess}. 
% Note that the integration of human attention is only needed in training. During inference, no eye-tracking data is used. % CRFIX

Our approach is inspired by the following intuition:  attention switches between two token literals may represent a synthesis of information between them~\cite{ahmad2023we, ali2015empirical}, 
% so attention switches between two token literals 
which may also signify a relation between these tokens during human code comprehension. 
Therefore, we aim to incentivize 
the self-attention mechanism within Transformers to discern the implicit relatedness that attention switches reveal between input tokens. 
As we discuss in Section~\ref{sec:prelim:transformer}, the self-attention naturally associates tokens with similar embeddings. Thus, we facilitate the recognition of relatedness between token pairs that are connected by an attention switch by introducing a shared embedding component for such pairs. Following this intuition, the detailed process of modeling human attention into Transformers can be divided into two steps: (1) mapping AST tokens and attention switches onto distinct embedding spaces and (2) combining the distinct embeddings obtained in the previous step to create a single input embedding vector for each token.

\textbf{Mapping Data to Three Embedding Spaces}
For each AST, we perform a breadth-first-search traversal starting at the root node to obtain a sequential representation of the AST, as illustrated in Figure~\ref{fig:model} (c). We denote this sequence of tokens $\{a, b, c, \dots\}$.  In standard Transformer architectures, each token is depicted by a semantic embedding and a positional embedding:
\begin{align*}
\text{Semantic Embedding } &\{E_a, E_b, E_c, \dots\} := \{f_e(a), f_e(b), f_e(c), \dots\}, \\
\text{Positional Embedding } &\{H_a, H_b, H_c, \dots\} := \{f_h(h(a)), f_h(h(b)), f_h(h(c)), \dots\}.
\end{align*}
Here, $h$ is a function that returns the height of the input token on the AST, and $f_e, f_h$ map to distinct embedding spaces. Intuitively, the semantic embedding vector captures the token's \textit{type}, and the positional embedding vector represents the token's \textit{height} on the AST. % ZK - height as in the tree depth? Eric: I'll change this to tree depth and change the diagrams too

After projecting AST tokens into these two embedding spaces, we proceed to convert attention switches into their own embedding representation. Each attention switch can be represented as a directed edge linking two AST nodes, as illustrated in Figure~\ref{fig:model} (b). Formally, a sequence of attention switch edges on an AST is denoted as $\{s^1_{\alpha_1\beta_1},s^2_{\alpha_2\beta_2}, s^3_{\alpha_3\beta_3}, \dots\}$. The superscripts align the attention switch edges in temporal order, and $\alpha_i, \beta_i\in\{a, b, c, \dots\}$ % ZK - should a, b, c here be 1, 2, 3 instead? Based on subscripts in the formal definition of attention switches
denote the origin and the target tokens of the directed edge. For example, $s^2_{ce}$ (with superscript $2$, $\alpha_2=c, \beta_2=e$)  denotes a participant's second attention switch in a given AST, transitioning from token `c' to `e'.  We use an embedding mapping $f_p$ to obtain embedding representations of attention switch edges:
$$\text{Attention Switch Embedding }\{P^1_{\alpha_1\beta_1},P^2_{\alpha_2\beta_2}, P^3_{\alpha_3\beta_3}, \dots\} : = \{f_p(s^1), f_p(s^2), f_p(s^3), \dots\}$$
Note that $f_e, f_h, f_p$ are distinct functions mapping to three separate embedding spaces. Furthermore, $f_p$ disregards the origin and target tokens of an attention switch edge. Hence, attention switch edges across different Java methods possessing identical superscripts (temporal cardinality) receive the same embedding representation. For instance, if two attention switch edges in two different Java methods are both the earliest attention switch for their respective methods, they are assigned the same embedding vector. 
In doing so, we generalize attention switch embeddings across different methods and encode temporal ordinality information into attention switch embeddings. % ZK suggests putting this reasoning right after the "Furthermore, f_p disregards..." sentence to give some motivation for the rest of the paragraph. I was having trouble following the importance of this business with f_p

\begin{figure}[tb]
\centering
\includegraphics[width=.88\textwidth]{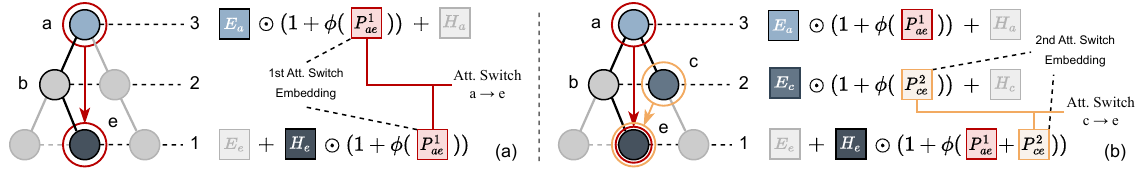}
\caption{Illustration of how \TheName{} models attention switch. Attention switches, conceptualized as edges connecting AST nodes, are mapped onto corresponding AST tokens by merging their respective embeddings.}
\label{fig:model_detail}
\end{figure}

\textbf{Combining Three Embedding Spaces}
Following these steps, we obtain sequences of semantic and positional embeddings of AST tokens via BFS on the AST graph. Additionally, the attention switch embeddings were derived as edges on a graph representation of the AST.
% ZK - may need an explanation why graph here is key, and the importance of a sequential representation
In this step, we map the attention switch edges onto corresponding AST tokens' embeddings, combining different embedding representations into a singular, sequential input embedding vector per AST token. We illustrate this mapping via two equivalent perspectives: explaining (i) what happens to each token and (ii) what happens to each attention switch edge.

\emph{Perspective (i): what happens to each AST token?} We answer this question with formal equations. Given some AST node, denoted $\gamma\in\{a, b, c, \dots\}$ with token encoding $E_\gamma$ and positional encoding $H_\gamma$. 
% ZK - reason/purpose for sum here? 
% kinda explained in perspective (ii) - maybe i should move up perspective (ii)? 
% Ah gotcha. Ok! 
We sum up the embeddings of all attention switch edges that originate from $\gamma$, denoted $\sum_{\alpha_k = \gamma} P^k_{\alpha_k\beta_k}$. Informally, this summation embodies all attention switches originating from token $\gamma$. We then map this sum onto the $\gamma$'s semantic embedding $E_\gamma$ through 
\begin{equation}\label{eq:E}
    E_\gamma\odot(1+\phi(\sum_{\alpha_k = \gamma} P^k_{\alpha_k\beta_k})),
\end{equation} where $\odot$ is element-wise multiplication and $\phi$ is the \emph{ReLU} function. Next, with a similar process, we sum up the attention switch embeddings of all attention switch edges that point to $\gamma$, denoted $\sum_{\beta_k = \gamma} P^k_{\alpha_k\beta_k}$. We then map this value onto the $\gamma$'s positional encoding $H_\gamma$, obtaining 
\begin{equation}\label{eq:H}
    H_\gamma\odot(1+\phi(\sum_{\beta_k = \gamma} P^k_{\alpha_k\beta_k})).
\end{equation}
 Finally, we sum up Expressions~(\ref{eq:E}) and (\ref{eq:H}), which are $\gamma$'s semantic and positional embeddings, both modified by attention switch embeddings, to obtain the final embedding representation of $\gamma$:
\begin{equation}\label{eq:E+H}
        E_\gamma\odot(1+\phi(\sum_{\alpha_k = \gamma} P^k_{\alpha_k\beta_k})) + H_\gamma\odot(1+\phi(\sum_{\beta_k = \gamma} P^k_{\alpha_k\beta_k}))
\end{equation}
For token $\gamma$, Expression (\ref{eq:E+H}) captures the token's semantic and positional embeddings, as well as accounting for all attention switch edges originating from and leading to token $\gamma$. Expression (\ref{eq:E+H}) is then used as token $\gamma$'s final input embedding vector fed into the Transformer. 

% \begin{figure}[tb]
% \centering
% \includegraphics[width=.88\textwidth]{figs/model_detail.pdf}
% \caption{Illustration of how \TheName{} models attention switch. Attention switches, conceptualized as edges connecting AST nodes, are mapped onto corresponding AST tokens by merging their respective embeddings.}
% \label{fig:model_detail}
% \end{figure}

\emph{Perspective (ii): what happens to each attention switch edge?} We answer this question through an exemplary case study, illustrated in Figure~\ref{fig:model_detail} (a). The attention switch edge described in the figure, being the first attention switch from a to e, can be denoted $s^1_{ae}$ with embedding $P^1_{ae}$. We map this edge's embedding $P^1_{ae}$ onto token $a$'s semantic embedding $E_a$, obtaining $E_a\odot(1+\phi(P^1_{ae}))$. Then, we map the same edge's embedding onto token $e$'s positional embedding $H_e$, obtaining $H_e\odot(1+\phi(P^1_{ae}))$. In this way, the direction of this edge is preserved: the mapping onto the \textit{input} embedding indicates $a$ is the origin of the attention switch edge, while the mapping onto \textit{positional} embedding indicates $e$ is the target of the edge. 

Note that in Figure~\ref{fig:model_detail} (a), after we add up each token's semantic and positional embeddings, both modified by attention switches, the resulting embedding representations of token $a$ and $c$ become more related.
This is because both token $a$ and $e$'s aggregated embedding vectors now share a same embedding component: $P^1_{ae}$.
As illustrated in this example, our method introduces relatedness through a shared embedding component between token pairs connected by attention switch. Intuitively, such relatedness is recognizable by self-attention mechanisms due to the mechanism's proclivity to associate tokens with similar embeddings.
Figure~\ref{fig:model_detail} (b) illustrates a similar process when a second attention switch edge is mapped onto the tokens, and this iterative process continues until all attention switches are incorporated.

Perspective (i) and (ii) are two ways of looking at the exact same procedure. 
Regardless of the perspective, the attention switch embeddings are mapped onto AST tokens' semantic and positional embeddings, resulting in a unified input embedding vector for each AST token. This unified embedding is then used as input for the Transformer-based models, enhancing their performance in code summarization tasks. Since the choice and details of Transformer models depend on the requirements of the downstream summarization tasks, we will introduce our Transformer architecture design in Section~\ref{sec:experiment}.

In essence, our approach allows us to implicitly represent directed human attention switch edges using Transformer's standard input embedding modality. Consequently, human attention switches during code comprehension 
%informally 
establish connections between AST tokens in a format that is intuitively compatible and recognizable by Transformers' self-attention layers.

% In general, mapping attention switch embeddings onto the AST provides incentives for the Transformer to deem two tokens more relevant if they are connected by a scanpath. Consequently, we nudge the Transformer to inter-relate pairs of tokens that are associated by human programmers.

% [move perspective 2 above perspective 1?]

\section{Experimental Design}\label{sec:experiment}

In this section, we present the experimental design for evaluating \TheName{}. Specifically, 
%we outline four research questions, 
we detail (1) dataset preparation, (2) model design and choices of hyperparameters, (3) noise and dropout training designs to mitigate bias in the eye-tracking dataset, and (4)  evaluation metrics.  We then use this design to answer four Research Questions described in Section~\ref{sec:analysis}.

\subsection{Dataset Preparation}

We briefly introduce the preparation of our dataset, encompassing both the collection of the eye-tracking dataset and the subsequent tokenization of the train-test dataset.

\begin{figure}[t]
\centering
\includegraphics[width=.9\textwidth]{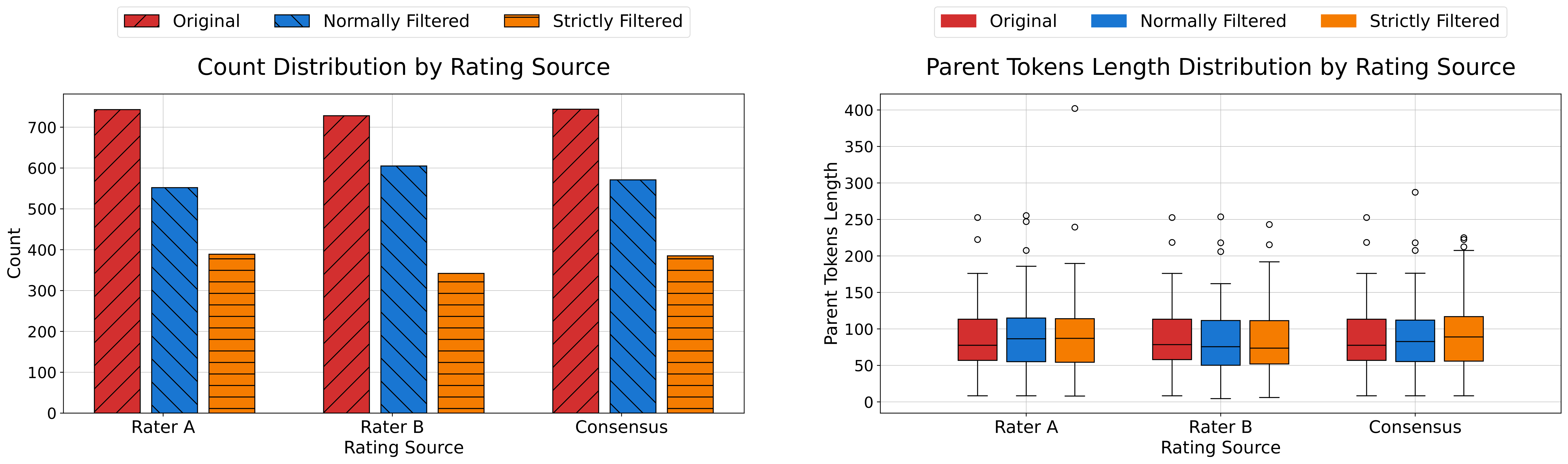}
\caption{Count and Parent Token Length Distribution by Rating Source. The count decreases for all rating sources as filtering tightens. The consensus approach mitigates outliers, which are seen in strictly filtered comments from Rater A.}
\label{fig:rating_comparison}
\end{figure}

% \begin{figure}[t]
% \centering
% \includegraphics[width=.92\textwidth]{figs/comparison_curve.pdf}
% \caption{Illustration of distribution curves for three data sources. For both Rater A and Rater B, there is bias in rating the comments, such as a preference for longer comments (Rater A) and not being strict enough in rating (Rater B), and consensus mitigates the bias.}
% \label{fig:curve_comparison}
% \end{figure}

\textbf{Eye-Tracking Dataset} We obtained the raw eye-tracking data from the human study as introduced in Section~\ref{sec:eye-tracking}, along with the summaries participants wrote. 
An inaccurate summary indicates lesser comprehension of the code, so we excluded eye-tracking data associated with inadequate summaries.
Specifically, we conducted manual filtering by two raters, Rater A and Rater B, to control for data quality. Each rater first assessed the written summaries independently using four metrics: Accuracy $(a)$, Completeness $(b)$, Verbosity $(c)$, and English Proficiency $(d)$, each rated from 1 to 5. Subsequently, the raters met in person to reconcile any discrepancies in their scores and reach a consensus. As a result, we established three filtering standards based on these metrics. For the original standard (Original), we retained all valid data points without considering their scores; for the filtered dataset (Filtered), we included data with $a \geq 3, d \geq 3$ and $b \leq 3, d \leq 3$; for the strictly filtered dataset (Strict), we used data with $a > 3, d > 3$ and $b < 3, d < 3$. Figure~\ref{fig:rating_comparison} displays the count and parent token length distribution by rating source. Through filtering, we introduced three levels to data quality, thereby facilitating the analysis of how different levels of eye-tracking data quality impact the performance of Transformer models. We chose the consensus data for train-test data preparation.

% ZK - is there a way to back up this last sentence?

\textbf{Train-Test Dataset} We adhere to the approach introduced in Sections~\ref{src_code_preprocess}~and~\ref{sec:preprocess} to tokenize the data and formulate the initial dataset. For the experiments, we eliminate all duplicate augmentations of ASTs to ensure that each combination of AST and eye-tracking data is unique. The final dataset comprises 10,676 original, 7,949 filtered, and 4,991 strictly filtered ASTs, each combined with eye-tracking data. For the Functional Summarization task, each augmented AST retains the original label of the source code. For the labels of the General Summarization task, we leverage an LLM~\cite{peng2023instruction} to paraphrase comments into semantically equivalent but varied comments to avoid reducing the task to mere memorization of comment sequences.  We follow the typical approach of allocating 80\% of the data to the training set and 20\% to the test set, and set the random seed to 42 during dataset preparation. 
% The test set contains no eye-tracking data. %CRFIX
Before being fed into the Transformer models, initial AST and eye-tracking datasets are transformed into embedding representations, as described in Section~\ref{approach:modeling}.

%[we also need to briefly mention the eye-tracking dataset here].

\subsection{Transformer Models and Hyper-Parameters}~\label{sec:experiment_transformer}

We now present our choices of Transformer models that consume the embedding representations of input datasets (obtained through methods described in Section~\ref{sec:approach}) and perform the downstream summarization tasks. 
We detail the structures used for both Functional Summarization and General Code Summarization, as well as our choices of hyperparameters. 
For Functional Summarization, we implement 4 encoder layers and use the \emph{CLS} token as the aggregated embedding for function classification.
For General Code Summarization, we employ a seq2seq Transformer structure, composed of 4 encoder layers and 4 decoder layers. The embeddings from the decoder layers are used as inputs for $Q$ and $K$ after the encoder layers in the encoder-decoder attention mechanism. For both model types, we incorporate default residual connections~\cite{he2016deep} and layer normalization~\cite{ba2016layer} in the feed-forward layers.

% \textbf{General Code Summarization} 4x Encoder and 4x Decoder, each has ;

% \textbf{Extreme Code Summarization} We use 4x Transformer Encoder, , max_number=300, max_height=50, num_classes=300

\textbf{Hyper-Parameters} For both the Functional Summarization and General Code Summarization models, we configure the total number of training epochs to 10, the learning rate to 1e-3, the hidden dimension to 32, and the number of attention heads to 4, employing Adam~\cite{kingma2014adam} with default settings as the optimizer. Unless specifically mentioned, we establish the maximum token of AST to 200, the maximum height to 50, the maximum number of classes to 300, the maximum number of vocab for comments to 2000, and the maximum length of summarization to 30. Additionally, we set the batch size to 256 during training and choose random seeds of 0, 1, 42, 123, and 12345 to repeat our experiments five times to obtain averaged results.

\subsection{Dropout and Noise in Training Design}

In addition to the standard hyperparameters, we introduce token-level dropout and Gaussian noise to simulate variations in data, thus creating increasingly challenging summarization tasks. Within such training scenarios, models  need to not only perform well on the provided data but also be robust to in- and out-of-domain variations.

\textbf{Robustness} We evaluate the \TheName{} Transformer models' robustness by applying dropout to semantic token embeddings. To represent varying levels of adaptability, we define three dropout rates: 0.0, 0.1, and 0.5, denoted as $R_0$, $R_1$, and $R_2$ respectively. $R_0$ illustrates our model's performance under normal settings without dropout. We further assess model robustness by integrating Gaussian noise into semantic token embeddings. We introduce three levels of Gaussian noise: 0.0, 0.1, and 0.5, represented as $N_0$, $N_1$, and $N_2$ respectively, with $N_0$  demonstrating our model's performance in a standard setting without noise. Specifically, for RQ3, we employ an interval of 0.05 to generate a uniform mesh of noise levels between $N_1$ and $N_2$ for performance visualization. 

% The introduction of dropout and Gaussian noise are only present in our answer to RQ3 in Section~\ref{sec:rq3} and are removed everywhere else to accurately assess the model's performance.

\subsection{Evaluation Metrics} 

We present our chosen quantitative evaluation metrics for both Ceneral and Functional Summarization, respectively. 

For quantitative evaluation of Functional Summarization, we employ Mean Average F1 (MAF1), Mean Average Precision (MAP), and Mean Average Recall (MAR) to assess the model’s performance in recovering the name of single functions. Besides these metrics, we conduct a case study to qualitatively evaluate how the introduction of human attention brings changes to Transformer's self-attention layers.

For quantitative evaluation of General Code Summarization, we use several ROUGE scores~\cite{lin2004rouge}, comprising the unigram-based score, ROUGE-1; bigram-based scores, ROUGE-2, ROUGE-S, and ROUGE-SU; and the sentence-based score, ROUGE-L, to contrast the performance of \TheName{} and a baseline Transformer model. For each metric, we calculate the mean average to represent the general performance on single functions. We recognize that ROUGE scores might not be optimal~\cite{stapleton2020human}, as there could be more apt metrics for code summarization~\cite{shi2022evaluation}. Nonetheless, we opt for ROUGE because it aligns well with the intrinsic nature of learning-based models in varied domains and eases comparative analysis for researchers. 

% \subsection{Hardware Settings}
% In our experiments, we use 

\section{Experimental Results}\label{sec:analysis}

%In this section, we present our results in the following sequence: For RQ1, we characterize programmers' gaze patterns to demonstrate that the incorporation of human attention is rational and has the potential to enhance machine learning models. In RQ2, we perform a quantitative comparison of our model with the Transformer across two tasks and extend the comparison to learning efficiency, generalizability, and robustness in RQ3. For RQ4, we visualize the attention maps from both models to qualitatively present the alterations in machine attention brought about by the incorporation of human attention. The detailed results are as follows:
% EL: feel like we need to make a BIG disclaimer here saying that the robustness and generalizability we claim are strictly the AI community's notion of robustness. We need to make surre that the SE reviewers do not see this keyword "Generalbility" as claiming that we can train on this dataset, and then test on other datasets WITHOUT human eye-tracking data.

In this section, we answer four research questions based on our experimental design to demonstrate the effectiveness of integrating eye-tracking data in training Transformer models for neural code summarization:

%We formulate the following research questions to demonstrate the efficacy of eye-tracking data in training Transformer models for neural code summarization:

\squishlist
    \item RQ1: What attention patterns do programmers exhibit as they read code? Are these patterns significant enough to be incorporated into machine learning models?
    \item RQ2: By integrating human and machine attention, to what extent can the performance of the Transformer model be enhanced?
    \item RQ3: How does incorporating human attention into the Transformer model improve learning efficiency and robustness?
    \item RQ4: What specifically changes in the self-attention layers of the Transformer when it is combined with human attention?
    % EL: since I didn't introduce our transformer architecture in the Approach section, we may need to briefly introduce it here.
    % \item RQ3: learning speed, generalizability and robustness
    %func name recovery, summarization, how attention map changes
\squishend

%We discuss each Research Question in turn. 

%We first discuss how we prepared out dataset incorporating the collection of eye-tracking data and source code summaries.  Then, we answer each research question in turn.
%To address RQ1, we present findings from our examination of the eye-tracking dataset. For RQ2 and RQ3, we devise two code summarization tasks: \textit{General Code Summarization}~(seq2seq text generation) and \textit{Functional Summarization}~(functionality classification), as introduced in Section~\ref{sec:intro}. The design details are as follows:
% EL: What about RQ3,4? 

% \begin{figure}[tb]
% \centering
% \includegraphics[width=.64\textwidth]{figs/token_tuple_distributions.pdf}
% \caption{Illustration of tuple simplification. By removing duplicate attention switching, the histograms of unique tuple lengths are more representative of data characteristics than those of recombined token lengths.}
% \label{fig:token_tuple_distribution}
% \end{figure}

\subsection{RQ1: Human Attention}
To determine whether human attention is suitable for integrating with a Transformer model, we first needed to characterize programmers' gaze patterns. Informally, we sought to determine whether code reading patterns are consistent and meaningful across programmers.
We first considered the amount of time participants spent reading the code, as well as their fixations, then calculated the most common attention switches (cf. Section~\ref{sec:preprocess}).

% ZK thinks this paragraph can be deleted if we're short on space
We find that participants looked at each Java method for an average of 26.54 seconds ($\sigma$ = 23.16s). Next, we find that participants averaged 94.92 fixation counts on each method ($\sigma$ = 43.81 fixations). 
Here we consider the \textit{average} fixation durations to describe the individual characteristics of each fixation, as opposed to a cumulative measure of fixation durations~\cite{sharafi2020practical}. 
During the Java summarization tasks, programmers' average fixation durations were 0.114 seconds ($\sigma$ = 0.037s).
Based on the standard deviations in particular, it appears programmers show variety in their behavior on the task. 
%However, these aggregate metrics do not reveal any specifics related to coders' reading behavior. 

Code comprehension research suggests that programmers do not read code linearly, instead revisiting certain elements~\cite{busjahn2015eye}, and focusing their attention on a subset of the code~\cite{crosby2002roles}.
To uncover whether predominant code reading patterns are present, we calculate the most common attention switches programmers made. 
Specifically, we first compiled the ordered sequences of programmers' fixations (i.e., scan paths). These lists provide an ordered record of the token literals that participants read. For instance, a programmer may read code in the following order: \texttt{String s}
$\xrightarrow{}$ \texttt{``hello world''} $\xrightarrow{}$ \texttt{return s}. This sequence is comprised of token literals, but in our analyses, we replaced literals with their semantic categories. The example above would become \texttt{variable declaration} $\xrightarrow{}$ \texttt{literal} $\xrightarrow{}$ \texttt{return}. 
Within these sequences of semantic categories, we then computed the most common pairs, such as \texttt{variable declaration $\xrightarrow{}$ literal}, and \texttt{literal $\xrightarrow{}$ return} from above. Within our dataset, we collected these ordered sequences of token literals for each participant, for every method they summarized, totaling 60,411 data points. 

Calculating the most common pairs (i.e., attention switches) in this data, we find that programmers most frequently look from \texttt{method declaration} $\xrightarrow{}$ \texttt{variable declaration} (2,593). Next, we see the reverse: \texttt{variable declaration} $\xrightarrow{}$ \texttt{method declaration} (2,533). The third (2,189) and fourth (2,179) most common attention switches follow this reversing pattern, where programmers most commonly vacillate between the same two semantic categories: \texttt{conditional statement} $\rightleftarrows{}$ \texttt{loop body}. This persists for the fifth (1,615) and sixth (1,588) most common attention switches as well: \texttt{conditional statement} $\rightleftarrows{}$ \texttt{method declaration}. 
By analyzing the most common attention switches made by participants, we see structured code reading patterns emerge, where programmers vacillate between categories (i.e., \texttt{loop body} $\rightleftarrows{}$ \texttt{conditional statement}), and focus comparatively more on conditional statements. These results illustrate patterns in how humans read code, which may be beneficial for code comprehension. Thus, we next investigate the impact of integrating these attention patterns into Transformer models.

% both complements and enhances previous findings into how programmers revisit parts of code, and what code reading strategies they use. 

% FIXED YU(?): [land on something like: there is indeed pattern in how humans read code - such patterns may be beneficial for code comprehension, thus we investigated integration this pattern into ML models.]

\subsection{RQ2: Quantitative Analysis}

\begin{table*}[t]
    \centering
    {
    \caption{Comparison of \TheName{} against Transformer on Functional Summarization. In the table, $R_1$ and $R_2$ represent dropout rates of 0.1 and 0.5, respectively, while $N_1$ and $N_2$ represent Gaussian noise levels of 0.1 and 0.5, respectively. The average value of each data point was determined by running the experiments five times, using 0, 1, 42, 123, and 12345 as random seeds.
    \label{tab:rq2_extreme}}
    \scalebox{0.58}{\begin{tabular}{l | r r r r | r r r r | r r r r}
        \toprule
        & \multicolumn{4}{c|}{MAF1@1} & \multicolumn{4}{c}{MAP@1} & \multicolumn{4}{c}{MAR@1} \\
        \midrule
        Metrics & $(R_1, N_1)$ & $(R_2, N_1)$ & $(R_1, N_2)$ & $(R_2, N_2)$ & $(R_1, N_1)$ & $(R_2, N_1)$ & $(R_1, N_2)$ & $(R_2, N_2)$ & $(R_1, N_1)$ & $(R_2, N_1)$ & $(R_1, N_2)$ & $(R_2, N_2)$ \\
        \midrule
        Transformer~(Original) 
        & 96.90 & 64.62 & 90.47 & 49.70 
        & 96.53 & 61.90 & 88.46 & 46.85 
        & 97.74 & 71.29 & 92.90 & 57.74 \\
        \TheName~(Original) 
        & 99.61 & 70.31 & 93.10 & 56.43 
        & 99.56 & 68.14 & 92.26 & 53.85 
        & 99.68 & 76.13 & 94.84 & 63.55 \\
        \midrule
        Improvement
        & +2.80\% & +8.79\% & +2.91\% & \textbf{+13.52\%}
        & +3.15\% & +10.11\% & +4.29\% & \textbf{+14.95\%}
        & +1.98\% & +6.80\% & +2.09\% & \textbf{+10.06\%} \\
        \midrule
        \midrule
        Transformer~(Filtered) 
        & 92.78 & 53.94 & 75.78 & 42.59 
        & 91.90 & 51.58 & 73.67 & 39.99 
        & 94.47 & 60.43 & 80.43 & 50.21 \\
        \TheName~(Filtered) 
        & 96.09 & 58.44 & 89.74 & 54.40 
        & 95.61 & 56.01 & 88.74 & 51.95 
        & 97.02 & 65.11 & 91.92 & 61.28 \\
        \midrule
        Improvement
        & +3.56\% & +8.35\% & +18.42\% & \textbf{+27.82\%}
        & +4.03\% & +8.59\% & +20.51\% & \textbf{+29.91\%}
        & +2.71\% & +7.73\% & +14.33\% & \textbf{+22.03\%} \\
        \midrule
        \midrule
        Transformer~(Strict) 
        & 82.92 & 52.76 & 76.54 & 46.70 
        & 81.36 & 49.21 & 74.11 & 42.95 
        & 86.45 & 61.29 & 81.94 & 55.48 \\
        \TheName~(Strict) 
        & 95.68 & 55.58 & 83.87 & 49.48 
        & 95.15 & 52.15 & 82.32 & 45.71 
        & 96.77 & 63.87 & 87.10 & 58.71 \\
        \midrule
        Improvement
        & \textbf{+15.38\%} & +5.33\% & +9.58\% & +5.96\%
        & \textbf{+16.94\%} & +5.97\% & +11.05\% & +6.43\%
        & \textbf{+11.95\%} & +4.21\% & +6.30\% & +5.80\% \\
        \bottomrule
    \end{tabular}}
    }
    %\vspace*{2mm}
\end{table*}\label{tab:rq2_results_extreme}

\begin{table*}[t]
    \centering
    \setlength{\tabcolsep}{1mm}{
    \caption{Comparison of \TheName{} against Transformer on General Code Summarization. In the table, $R_0$ and $R_1$ represent dropout rates of 0.0 and 0.1, respectively, while $N_0$ and $N_1$ represent Gaussian noise levels of 0.0 and 0.1, respectively. The average value of each data point was determined by running the experiments five times, using 0, 1, 42, 123, and 12345 as random seeds.
    \label{tab:rq2_general}}
    \scalebox{0.64}{\begin{tabular}{l | r r | r r | r r | r r | r r }
        \toprule
        & \multicolumn{2}{c|}{ROUGE-1} & \multicolumn{2}{c|}{ROUGE-2} & \multicolumn{2}{c|}{ROUGE-S} & \multicolumn{2}{c|}{ROUGE-SU} & \multicolumn{2}{c}{ROUGE-L} \\
        \midrule
        Metrics & $(R_0, N_0)$ & $(R_1, N_1)$ & $(R_0, N_0)$ & $(R_1, N_1)$ & $(R_0, N_0)$ & $(R_1, N_1)$ & $(R_0, N_0)$ & $(R_1, N_1)$ & $(R_0, N_0)$ & $(R_1, N_1)$\\
        \midrule
        Transformer~(Original) & 71.55 & 64.89 & 49.30 & 40.29 & 45.35 & 36.51 & 52.63 & 44.39 & 69.52 & 62.75 \\
        \TheName~(Original) & 72.91 & 66.92 & 50.67 & 42.76 & 46.79 & 38.84 & 53.95 & 46.47 & 70.82 & 64.85 \\
        \midrule
        Improvement & $+1.90\%$ & $\textbf{+3.13\%}$ & $+2.78\%$ & $\textbf{+6.12\%}$ & $+3.17\%$ & $\textbf{+6.39\%}$ & $+2.51\%$ & $\textbf{+4.70\%}$ & $+1.87\%$ & $\textbf{+3.35\%}$ \\
        \midrule
        \midrule
        Transformer~(Filtered) & 64.90 & 58.45 & 40.02 & 33.03 & 35.73 & 28.37 & 43.77 & 36.61 & 62.98 & 56.47 \\
        \TheName~(Filtered) & 66.89 & 60.18 & 42.01 & 34.44 & 37.71 & 29.79 & 45.63 & 38.01 & 64.91 & 58.16 \\
        \midrule
        Improvement & $\textbf{+3.07\%}$ & $+2.96\%$ & $\textbf{+4.97\%}$ & $+4.27\%$ & $\textbf{+5.54\%}$ & $+5.00\%$ & $\textbf{+4.25\%}$ & $+3.82\%$ & $\textbf{+3.05\%}$ & $+3.00\%$ \\
        \midrule
        \midrule
        Transformer~(Strict) & 48.18 & 44.58 & 24.41 & 22.30 & 16.26 & 13.03 & 24.27 & 20.74 & 46.67 & 43.30 \\
        \TheName~(Strict) & 48.45 & 44.92 & 24.36 & 22.28 & 16.51 & 13.24 & 24.53 & 20.95 & 46.99 & 43.64 \\
        \midrule
        Improvement & $+0.56\%$ & $\textbf{+0.76\%}$ & $-0.21\%$ & $-0.09\%$ & $+1.54\%$ & $\textbf{+1.61\%}$ & $\textbf{+1.07\%}$ & $+1.01\%$ & $+0.68\%$ & $\textbf{+0.79\%}$ \\
        \bottomrule
    \end{tabular}}
    }
\end{table*}\label{tab:rq2_results_general}

%After examining the eye-tracking dataset to determine its appropriateness for use in machine learning, we begin by 
Having conjectured the usefulness of our eye-tracking dataset, we next quantitatively analyze \TheName{}'s demonstrated performance gain on neural code summarization tasks, focusing on both Functional Summarization and General Code Summarization tasks.

For Functional Summarization, Table~\ref{tab:rq2_extreme} illustrates \TheName{}'s substantially improved performance in MAF1, MAP, and MAR compared to the vanilla Transformer. We highlight the best performance across different scenarios and found that the improvement in MAF1 can be as high as 29.91\% (for MAP@1 under $R_2$ and $N_2$), showcasing promising potential. For the strictly filtered dataset, \TheName{} demonstrates improved performance on $R_1$ and $N_1$. However, this improvement declines in more challenging training scenarios with higher dropout and noise levels, indicating that training only with high-quality eye-tracking data may not be beneficial in all scenarios. For General Code Summarization, Table~\ref{tab:rq2_general} shows a consistent improvement across all ROUGE metrics for \TheName{}, except on the Strict dataset where the performances of \TheName{} and Transformer are comparable. We highlight the best performance across different regular training settings and found that on the Original dataset, \TheName{} generally performs better when there are regularization terms (i.e., 6.39\% for ROUGE-S under $R_1$ and $N_1$), while on the Filtered dataset, \TheName{} works better in a normal setting (i.e., 5.54\% for ROUGE-S under $R_0$ and $N_0$).

\subsection{RQ3: Robustness and Efficiency} \label{sec:rq3}

\begin{figure*}[t]
\centering
\includegraphics[width=0.94\textwidth]{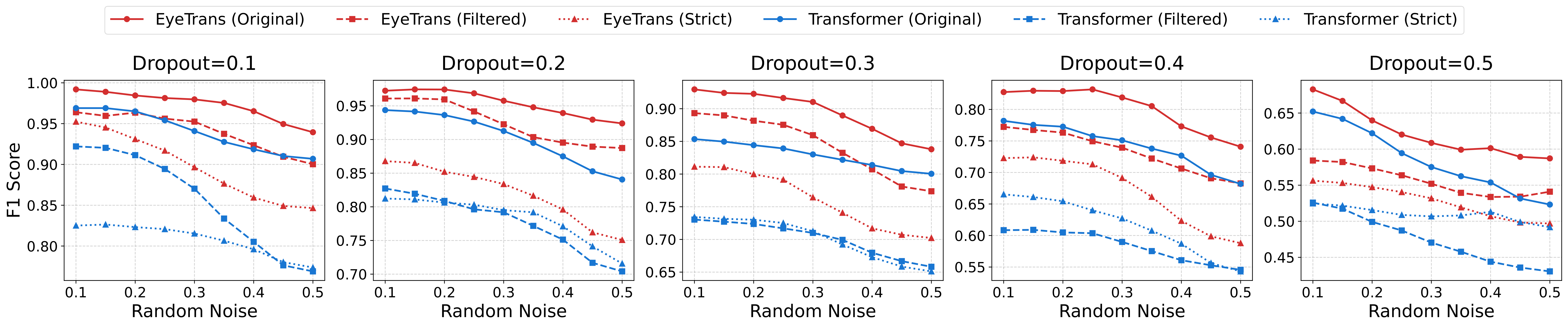}
\caption{Performance comparison of \TheName{} and Transformer on Functional Summarization with varying dropout and noise. The average MAF1 of each data point was determined by conducting the experiments five times, using 0, 1, 42, 123, and 12345 as random seeds.}
\label{fig:rq2_curves_extreme}
\end{figure*}

\begin{figure*}[t]
\centering
\includegraphics[width=0.94\textwidth]{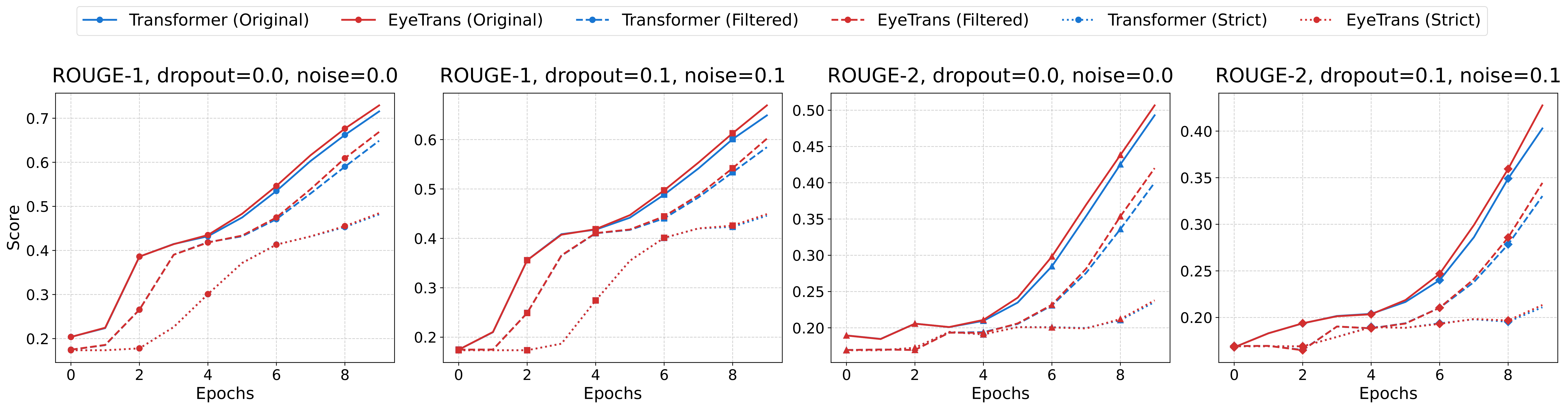}
\caption{Learning curve comparison of \TheName{} and Transformer on General Code Summarization with varying eye-tracking data quality. For simplicity, we use 10 epochs for each curve. The average ROUGE of each data point was determined by conducting the experiments five times, using 0, 1, 42, 123, and 12345 as random seeds.}
\label{fig:rq2_curves_general}
\end{figure*}

Based on RQ2, we know that by combining human attention, the performance of the Transformer has improved for both tasks. 
We further investigate two key aspects: (1) whether the performance changes under challenging training scenarios with noise and dropout, and (2) how training efficiency changes with respect to the quality of human attention data. 
%However, we may need further analysis to answer two questions: 

%\begin{enumerate}
%    \item How does performance change under different simulated training scenarios?
%    \item What is the trend in efficiency gains achieved by incorporating different quality human attention?
%\end{enumerate}

% FIXED: EL: again, the use of "real-world" here really seems to set the flag here for SE reviewers to assume that we can train on this dataset, and test on any raw dataset (without eye-tracking) in the wild -> changed to simulated real-world scenarios

This involves a detailed analysis of the robustness and training efficiency of \TheName{} with respect to data quality. To address (1), we plot the MAF1 curve for various combinations of human attention data qualities, dropout rates, and Gaussian noise levels, as depicted in Figure~\ref{fig:rq2_curves_extreme}.  The Figure shows dropout rates from 0.1 to 0.5 with intervals of 0.1 from left to right. In each subplot, the x-axis denotes the increase in Gaussian noise. 

% We notice that \TheName{} improves the robustness of Transformer in all subplots for the original, filtered, and strictly filtered dataset. At the same time, for the strictly filtered dataset, \TheName{} demonstrates improved performance on $R_1$ and $N_1$. However, this improvement declines in more challenging training scenarios, indicating that this filtering strategy is less robust.

\textbf{Greater Robustness} We note that robustness against dropout and noise is enhanced after integrating human attention with machine attention in \TheName{}. In Functional Summarization, \TheName{} exhibits improved performance overall when trained under increased difficulty (i.e., higher dropout rate and Gaussian noise), demonstrating greater robustness compared to the vanilla Transformer model, regardless of the human attention data quality.

For answering (2), we plot the ROUGE-1 and ROUGE-2 learning curves for each epoch during the General Code Summarization training process, as shown in Figure~\ref{fig:rq2_curves_general}. For comparison, we use only the first 10 epochs for each curve. In the Figure, we observe a steady improvement from \TheName{} in ROUGE-1 and ROUGE-2, particularly in the later stages of training with the Filtered dataset. For the strictly filtered dataset, due to the decrease in the training set size, both \TheName{} and Transformer struggle in the initial phase of training, making it challenging to distinguish \TheName{} from the baseline Transformer. This indicates it is critical to include both data quality and diversity when filtering human attention data during training.

In summarizing the characteristics of \TheName{} from both RQ2 and RQ3, we conclude: (1) the integration of eye-tracking data into the Transformer enhances the overall performance in code summarization tasks, improving the model's robustness, and (2) there is a pivotal equilibrium between data quality and diversity. The filtered dataset is advantageous for both RQs and generally outperforms the Original dataset, whereas the strictly filtered dataset sacrifices data diversity for quality, leading to reduced performance and training efficiency.

% \begin{figure*}[tb]
% \centering
% \includegraphics[width=0.98\textwidth]{figs/rq2_curves_general.pdf}
% \caption{ROUGE-1 \& ROUGE-2 Score Comparison of \TheName{} and Transformer on different settings. The average value of each data point was determined by conducting the experiments five times, using 0, 1, 42, 123, and 12345 as random seeds. The shades represent the standard deviation for each data point.}
% \label{fig:rq2_curves_general}
% \end{figure*}

% \begin{figure*}[tb]
% \centering
% \includegraphics[width=0.98\textwidth]{figs/rq2_curves_extreme_sharey.pdf}
% \caption{Mean Average F1 (MAF1) Score Comparison of \TheName{} and Transformer under different settings. The average value of each data point was determined by conducting the experiments five times, using 0, 1, 42, 123, and 12345 as random seeds. The shades represent the standard deviation for each data point.}
% \label{fig:rq2_curves_extreme}
% \end{figure*}

% Teacher Forcing During Training

\subsection{RQ4: Merged Attention Map} \label{sec:rq4}

\begin{figure*}[t]
\centering
\includegraphics[width=0.92\textwidth]{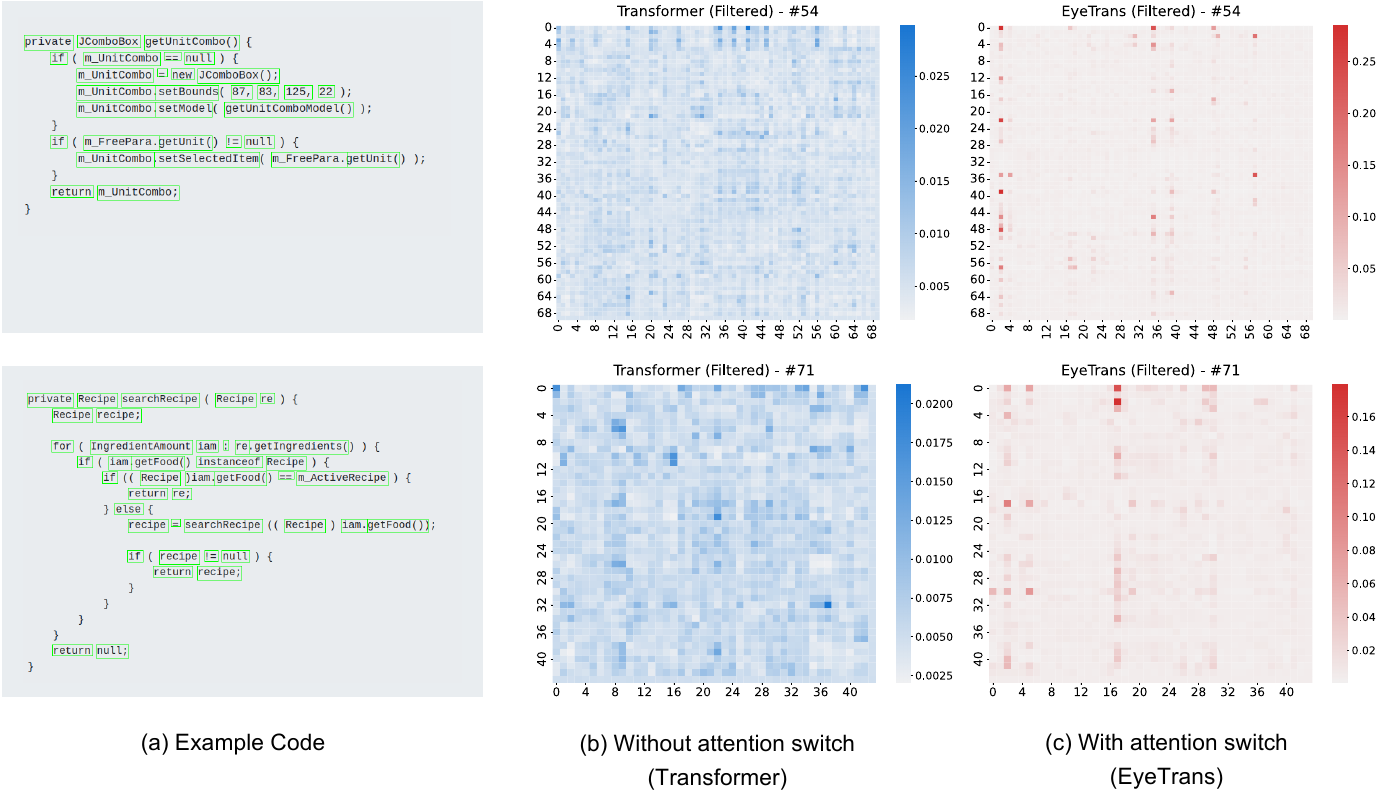}
\caption{Illustration of two examples on Attention Simplification in Functional Summarization. We use heatmaps to visualize the $QK^T$ matrix extracted from the first Transformer block in both models. (a) showcases the example Java code comprehended by the models (green boxes indicate the bounding boxes for eye-tracking data analysis); (b) visualizes the Transformer attention map without attention switch; and (c) visualizes the Transformer attention map with attention switch~(\TheName{}). We use models trained with a seed of 42 to ensure the reproducibility of the visualization maps.} % \yu{improve this figure. At least you should let readers know what each column and row is and what is the take home points from the illustrations}}
\label{fig:rq4_extreme}
\end{figure*}

In the preceding subsections, we examined the characteristics of eye-tracking data to substantiate our experiments and conducted a quantitative comparison between \TheName{} and Transformer concerning two code summarization tasks. In this subsection, we address the remaining question: specifically, what observable changes occur in the self-attention mechanism of the Transformer structure after incorporating human attention? To answer this question, we use visualizations of the $QK^T$ matrix in the first block of the Transformer as approximations of the model's attention~\cite{zeng2022extensive}. Specifically, as detailed in Section~\ref{sec:prelim:transformer}, the $QK^T$ matrix in Equation (\ref{eq:att}) conceptually captures the pair-wise inter-relation between input tokens. Thus, we refer to this matrix as \emph{attention map} and use heatmaps to portray it, effectively representing the Transformer’s attention patterns.
%A heatmap of the $QK^T$ matrix, as we use in Section~\ref{sec:rq4}, is thus an conceptual representation the inter-relation of input tokens deemed by self-attention. %why do we care about heat maps here?
After examining attention maps from both \TheName{} and a Transformer without human attention information, we identified one interesting pattern which we refer to as \textit{Attention Simplification}.

\textbf{Attention Simplification}  We observe that in both Functional Summarization and General Code Summarization, \TheName{} enhances the performance of the Transformer by refining the self-attention pattern. We use the attention map from Functional Summarization as examples for a case study, as shown in Figure~\ref{fig:rq4_extreme}. In the Figure, \TheName{} suppresses some unimportant relations in the matrix while emphasizing others, intensifying the effectiveness of selected patterns. This illustrates one potential mechanism where human attention provides pivotal guidance, eliminating redundant and useless weights to enhance the performance of the globally perceptive attention maps in the Transformer. Visualizing attention maps for both examples in the case study can reveal changes within the Transformer block caused by the incorporation of human attention, offering insights for future research.

The observed performance enhancement by \TheName{} and the identified refined attention pattern might also provide potential insights for future work, particularly in the domain of explainable AI, shedding light on how human attention can guide model refinement and facilitate the interpretation of model decisions. Furthermore, the implications of these findings might inspire more extensive explorations into the incorporation of human attention in model development, contributing to evolution of more intuitive, understandable, and reliable AI systems.

% the fact that the attention map changes can be viewed as generalizable, apply the finding to say it means this type of information to affect models in general, or the lesson here is SE in the future is increasingly rely on AI, and these AI models can learn much from human 

% sub sub section of rq4

% -> comment and attention map

% Exploratory analysis of eye-tracking data for machine learning -> thorough analysis of preprocessing -> 1 2 3 4 -> for each part, put a sentence indicating that this part is used later

\subsection{Ablation Studies}

% Besides the inherent ablation studies presented in our previous RQs, we further conducted ablation studies on four different experimental settings: (1) removing the height embedding from both models, (2) replacing the activation function $\phi$ from ReLU to Sigmoid, and (3) only use the $\phi(\sum_{\alpha_k = \gamma} P^k_{\alpha_k\beta_k})$ by removing the plus one. We use extreme code summarization with $R_2$, $N_2$ to examplify the changes. We observe that: about MAF1 for (1), both \TheName{} and Transformer drop by 12.74\% and 48.27\%, respectively, showing height embedding is a valid positional embedding for Transformer models during training; about MAF1 for (2) and (3), the performance of \TheName{} drops by 15.04\% and 32.76\%, respectively, demonstrating the effectiveness of our model design.

%In addition to the inherent ablation studies presented in our previous research questions (RQ2 and RQ3), 
We conduct further ablation studies under three distinct experimental settings:

\begin{enumerate}
    \item Removing the positional/height embedding term from both \TheName{} and Transformer.
    \item Replacing the activation function, \( \phi \), in Expression (\ref{eq:E+H}) from \emph{ReLU} to \emph{Sigmoid}.
    \item Substituting $E_\gamma\odot(\phi(\sum_{\alpha_k = \gamma} P^k_{\alpha_k\beta_k})) + H_\gamma\odot(\phi(\sum_{\beta_k = \gamma} P^k_{\alpha_k\beta_k}))$ in place of Expression (\ref{eq:E+H}), removing the $+1$ in the expression and thus reducing the relative importance of $E_\gamma$ and $H_\gamma$.
\end{enumerate}

We used Functional Summarization with \( R_2 \) and \( N_2 \) to exemplify the changes. Our observations are as follows: in ablation setting (1), both \TheName{} and the Transformer exhibit a drop by \(12.74\%\) and \(48.27\%\) in MAF1, respectively, indicating that height embeddings adequately encode positional information into Transformer models to improve performance. In settings (2) and (3), the performance of \TheName{} drops by \(15.04\%\) and \(32.76\%\) in MAF1, respectively, which provides justification for our model design choices. 
%implementation choices of our model design.

\section{Threats to Validity}\label{sec:validity}

There are two main threats to the validity of our evaluation. First, due to the specialized nature of human study research, we created a dataset with eye-tracking information to evaluate \TheName{}, which is based on Java. However, its effectiveness may vary with other programming languages. 
%Secondly, in analyzing \TheName{}, the Transformer structure inherits designs from \TheName{}, excluding attention-switching, which may not be optimal. % I don't understand this.
Second, for both the Functional Summarization and General Code Summarization tasks, the tasks and evaluation metrics used, such as ROUGE and MAF1, may not correlate with human developer performance. To minimize bias as much as possible, we selected multiple metrics and conducted several experiments, using fixed seeds.

\section{Related Work}\label{sec:background}

\TheName{} lies at the intersection of code summarization, human attention, and machine learning. In this section, we highlight the relevance of \TheName{} with past works in these domains.

\subsection{Code Summarization and Human Attention}

Eye-tracking has been used to measure human attention during various Software Engineering tasks~\cite{sharafi2020practical}, including code summarization~\cite{rodeghero2014improving, abid2019developer}. 
This technology is particularly suited to study programmers given its high accuracy~\cite{Tobii_2023}, and direct integration with developers' working environments~\cite{shaffer2015itrace}. Researchers have used eye-tracking to study debugging behaviors~\cite{p-Bednarik-strategy-2012}, differences between expert and novice coders~\cite{abid2019developer}, and code reading strategies~\cite{busjahn2015eye}, among others~\cite{Obaidellah:2018}. Within code summarization research, Rodeghero et al. conducted an experiment similar to the eye-tracking data collection in this study, where programmers' gaze was recorded as they wrote code summaries~\cite{rodeghero2014improving}. That work also aimed to improve methods for automated source code summarization by incorporating human attention. 
This approach did not use machine learning, instead selecting keywords for summaries based on where programmers fixated most. 
% By measuring programmers' fixations on the code, those researchers selected the  focused most.   

In the years since that paper was published, the advancement in deep learning propelled machine learning models to autonomously generate summaries for source code, a task referred to as neural code summarization. Since NeuralCodeSum~\cite{ahmad2020transformer} first introduced the use of Transformers in neural code summarization, many Transformer-based models have showcased remarkable performance across various task settings~\cite{wu2020code, shi2022evaluation, tang2022ast, gao2023code, gong2022source}. Notably, the leading approaches have significantly benefited from the structural information of source code, particularly by leveraging AST representations~\cite{alon2018code2seq, hu2018deep, leclair2020improved, shi2021cast, lin2021improving, tang2022ast, gong2022source, gao2023code}. \TheName{} elaborates upon this line of neural code summarization work by pioneering the integration of human attention data to improve model performance. 

\subsection{Integration of Eye-Tracking in Machine Learning} 

\TheName{} also builds upon past works that leverage eye-tracking to improve general machine-learning performance. 
In computer vision, several works have demonstrated improved performance by incorporating eye-tracking data~\cite{sugano2016seeing, qiao2018exploring, karessli2017gaze, xu2015gaze, yu2017supervising, fei2022attention}.
%The domain of computer vision prominently exemplifies the effectiveness of eye-tracking data in enhancing model outcomes~\cite{sugano2016seeing, qiao2018exploring, karessli2017gaze, xu2015gaze, yu2017supervising, fei2022attention}. 
Meanwhile, in NLP, the use of eye-tracking data has been shown beneficial in tasks such as syntactic labeling~\cite{klerke2019glance}, pronoun classification~\cite{yaneva2020classifying}, reference resolution~\cite{iida2011multi}, and multi-word expression prediction~\cite{rohanian2017using}. These studies exemplify the mainstream methodology in NLP for integrating eye-tracking data, primarily employing it as an additional set of input for the NLP models, distinct and separate from the original dataset.

Another prevalent methodology employs eye-tracking data as either a regularizing factor or an additional task to align neural networks' decision-making with human attention patterns~\cite{zhang2019using, barrett2018sequence, klerke2016improving}. Recently, a notable contribution from Sood et al.~\cite{sood2020improving} involved modifying the Luong attention layer within an LSTM, by introducing token-specific attention scores that mimic human eye fixations. 

\TheName{} advances former works by directly using Transformers' standard input embedding modality to represent human attention, eliminating past works' requirement to either modify the standard NLP model architectures or use eye-tracking data as a separate input set. Moreover, compared to past works' predominant focus on LSTM-centric networks, we pioneer the integration of human attention into modern Transformer-based architectures. We also pioneer the use of attention switches, rather than solely relying on fixation, as a representation of human attention to enhance the performance of machine learning models.

% First, most previous efforts predominantly revolved around LSTM-centric NLP networks~\cite{klerke2016improving, barrett2018sequence, zhang2019using, klerke2019glance, sood2020improving} and lacked compatibility with modern transformer architectures. Moreover, many of these models integrated human and machine attention by modifying standard NLP model architectures directly. In contrast, \TheName{} integrates human attention cues through a data-centric manner, infusing eye-tracking data into input embeddings. This decouples human attention cues from transformer architectures, avoiding direct architectural modifications and ensuring compatibility with larger industrial transformer models. (also just leveraged fixation, no attention switch)

% However, a gap remains as current models have yet to integrate human code comprehension patterns - that is, patterns indicating how humans read and understand code - to improve neural model performances. \TheName{} addresses this void, demonstrating that integrating human attention patterns with machine attention mechanisms can enhance the quality of code summarization.

% Most importantly, a gap previously existed in combining human and machine attention within NLP for code comprehension tasks, marking \TheName{} a pioneering stride in the domain.

% \textbf{TBD} TBD

\section{Future Works}\label{sec:future}

We consider \TheName{} as a first step towards integrating human and machine attention. As such, many potential extensions and ramifications of this work are yet to be explored. We discuss such future directions in this section.
% clearly stating that this is something yet to explore
% \section{Extensions to LLMs}

We intend EyeTrans to work in concert with Large Language Models (LLMs), not to compete with them. The key concepts of our paper is applicable to many Transformer-based models. The majority of LLMs today are fundamentally based on the Transformer backbone, but with more attention heads, a wider embedding vector size, scale increases, and other changes. We focus on improving the underlying Transformer architecture in this paper. The benefits we propose could in theory be applied to larger models, though we test them using small eye-tracking data and smaller models. This aligns with the current state of the art of eye-tracking studies in SE, which is often coupled with limitations on sample sizes. 

Yet, there may exist realistic approaches to overcome the costly acquisition of human visual attention data. For example, in reading natural languages (as compared to code), the E-Z Reader model~\cite{reichle2003ez} has been well-established in predicting human visual gaze. Applying E-Z-Reader-predicted pseudo-human-gaze on natural text, and consequently aligning NLP attention according to such pseudo-human-gaze, has improved model performance on natural languages~\cite{sood2020improving}. The SE community currently lacks well-established computational modeling of programmers’ gaze during code reading. However, steps towards this direction have recently been undertaken~\cite{bansal2023towards}. With future computational modeling techniques capable of adding accurate pseudo-visual attention to code, \TheName{} can be trained on regular, large datasets just like any other NLP model, without the need for extensive eye-tracking experiments.

\section{Conclusion}\label{sec:conclusion}
In this paper, we present \TheName{}, an approach to effectively incorporate human attention into the Transformer for neural code summarization tasks. In \TheName{}, human attention serves as a ``connection'' between two ends of the token embeddings, representing a data-centric incorporation of human attention without altering the Transformer structure. 
Integrating human attention in training results in a performance improvement of up to 29.91\% in Functional Summarization, and up to 6.39\% in General Code Summarization, thereby demonstrating the effectiveness of \TheName{}. 
This is the first, proof-of-concept work to integrate the eye-tracking patterns of programmers into Transformer models, achieving improvement on performance across two different code summarization tasks and comprehensive analysis.

We hope the basic concept of \TheName{} can be applied to numerous training scenarios that use the Transformer or Transformer block as a fundamental component. Moreover, with the future development of pseudo-eye-tracking paths and data augmentation on other structured and human-readable data in programming, our idea can be readily adopted. This will allow for the incorporation of a new modality during the training of the Transformer and has the potential to become a fundamental component of the Transformer. In the future, we plan to enhance \TheName{} by designing paired eye-tracking data augmentation methods, such as few-shot link prediction and label propagation, into the general model structure.

\section{Declarations}

In this section, we outline the compliance, data availability, funding sources, and acknowledgments associated with this study, highlighting the essential aspects that underpin our research.

\textbf{Ethical Compliance} This study was conducted in compliance with all applicable ethical standards and received approval from the Institutional Review Board (IRB). Informed consent was obtained from all individual participants involved in the study.

\textbf{Data-Availability Statement} All data and scripts are available on Zenodo~\cite{Zhang}. The repository includes comprehensive documentation to facilitate replication and extension of our research.

\textbf{Funding Statement} This research was supported by NSF CCF-2211429, NSF CCF-2211428, NSF CCF-2100035 and NSF SaTC-2312057.

\textbf{Acknowledgments} We appreciate the anonymous reviewers' feedback and constructive suggestions for improving our manuscript. We also greatly thank all the participants in our study for their participation.

\bibliographystyle{ACM-Reference-Format}
\bibliography{acmart}

\end{document}